\documentclass[final,3p,11pt,nopreprintline]{elsarticle}

\usepackage{amssymb}
\usepackage{amsmath}

\usepackage[final]{changes}
\usepackage{xcolor}
\definecolor{cud-vermilion}{HTML}{D55E00}
\definecolor{cud-blue}{HTML}{0072B2}
\definecolor{cud-bluish-green}{HTML}{009E73}
\definechangesauthor[name=r1, color=cud-vermilion]{r1}
\definechangesauthor[name=r2, color=cud-blue]{r2}
\definechangesauthor[name=r3, color=cud-bluish-green]{r3}

\usepackage{bm}
\usepackage{url}
\usepackage{hyperref}
\hypersetup{hidelinks}

\begin{document}

\begin{frontmatter}



\title{
  A spatial autocorrelation method for estimating Rayleigh-wave attenuation coefficients from microtremor array measurements
}


\author[isct]{Harusato Kimura}\ead{kimura.h.596d@m.isct.ac.jp} 
\author[isct]{Hitoshi Morikawa} 

\affiliation[isct]{
  organization={Department of Civil and Environmental Engineering, Institute of Science Tokyo},
  addressline={4259-G3-7, Nagatsuta, Midori-ku}, 
  city={Yokohama},
  postcode={226-8502},
  country={Japan}
}

\begin{abstract}
The velocity and attenuation structures of near-surface soils play an important role in controlling earthquake ground motions.
While numerous passive techniques have been developed for estimating velocity structure, practical methods for estimating attenuation from microtremor array measurements remain limited.
We developed a spatial autocorrelation (SPAC) method for estimating Rayleigh-wave attenuation coefficients using an attenuation-aware formulation of the SPAC coefficient.
A theoretical analysis of directional aliasing showed that attenuation estimation requires denser azimuthal sampling than conventional phase-velocity estimation, motivating the use of a regular heptagonal array.
The method uses a two-stage procedure in which initial estimates of phase velocity and the attenuation coefficient are obtained using a first-order approximation and then refined by fitting the complex SPAC coefficient.
Synthetic data analyses demonstrated that the proposed method can recover attenuation coefficients under ideal conditions and clarified how directional aliasing, record length and sensitivity to attenuation affect the estimates.
Comparison with a beamforming-based method showed that the proposed method is less sensitive to wavefield directionality and can be applied at longer wavelengths relative to the array size.
The method was further validated using microtremor-like waveforms generated by independent P--SV wave-propagation simulations.
Finally, a preliminary application to field microtremor records yielded attenuation coefficients that were broadly consistent across array sizes and with values expected from existing empirical attenuation relationships.
These results demonstrate the applicability of the proposed method to passive estimation of Rayleigh-wave attenuation.
\end{abstract}




\end{frontmatter}



\section{Introduction}\label{s:intro}

Accurate knowledge of the subsurface velocity and attenuation structure at a site is important for estimating earthquake ground motions.
Numerous methods have been proposed for estimating velocity structure, including active methods using artificial vibration sources~\cite{Park1999, Xia1999}, methods using earthquake observations~\cite{Kawase2011} and passive methods using microtremors~\cite{Aki1957, Capon1969, Horike1985, Arai2004, Lontsi2015}.
In contrast, fewer methods have been developed for estimating attenuation structure, and robust methods suitable for practical use are not yet well established.
In particular, passive methods for estimating local (i.e. near-surface) attenuation structure remain limited~\cite{Parolai2022}.

Previous efforts to estimate local attenuation structure passively have relied mainly on microtremor array measurements, following two broad approaches.
One approach extends seismic interferometry~\cite{Shapiro2005} and the spatial autocorrelation (SPAC) method~\cite{Aki1957, Cho2006a}, both of which are used to estimate velocity structure from microtremors~\cite{Albarello2009, Nakahara2012, Parolai2014, Boxberger2017, Cho2021a}.
The other, proposed by \citet{Abbas2025}, adapts an attenuation estimation method~\cite{Aimar2024a} developed for multichannel analysis of surface waves (MASW)~\cite{Park1999, Xia1999}, an active method for estimating near-surface velocity structure.
Although both approaches appear to have had some success, their theoretical formulations rely on substantially more restrictive wavefield assumptions than those conventionally used to estimate velocity structure from microtremor array measurements.

Most studies extending seismic interferometry and the SPAC method draw on either a method for estimating regional-scale attenuation structure using seismic interferometry~\cite{Prieto2009} or the theoretical framework developed from that method by \citet{Nakahara2012}.
These studies use array sizes, target wavelength ranges and observation durations typical of the SPAC method.
However, their theoretical justification is drawn from seismic interferometry on the basis of its similarity to the SPAC method~\cite{Nakahara2012}.
This creates a mismatch between an analysis procedure based on the SPAC method and wavefield assumptions drawn from seismic interferometry.
Specifically, the conventional SPAC method assumes a target wavefield consisting of a ``superposition of mutually uncorrelated Rayleigh waves arriving as plane waves from arbitrary directions with arbitrary intensities''~\cite{Cho2006a}, whereas this approach assumes a ``superposition of waves emitted by numerous point sources with a spatially homogeneous random distribution''.
Applying a method based on the latter assumption to microtremor array records collected under conditions where the former is widely accepted therefore requires appropriate justification.
Even with such justification, assuming a spatially homogeneous random source distribution imposes a substantial additional restriction on the wavefield.

The Abbas method, derived from MASW, interchanges the phase and amplitude-variation components of the observed records by raising the Fourier spectra to the power of the imaginary unit.
It then applies the frequency--wavenumber spectral method (FK method), conventionally used to estimate Rayleigh-wave phase velocities.
This simple but powerful approach enables stable estimation of Rayleigh-wave attenuation coefficients through appropriate spectral normalisation.
However, its origin in active-source surface-wave exploration imposes a major theoretical restriction: the target wavefield must be nearly unidirectional.
In practice, when microtremor records are divided into short segments, the wavefield must be unidirectional within each segment.
This still represents a substantial restriction relative to the conventional wavefield assumptions used in microtremor array surveys with the SPAC or FK method.

To address these limitations, \citet{Kimura2026a} incorporated attenuation effects directly into the SPAC formulation while retaining the conventional physical assumptions about the wavefield.
They derived the corresponding expression for the SPAC coefficient, a fundamental quantity in the SPAC method.
Using synthetic data for a specified subsurface structure, they then examined whether attenuation effects were discernible in the SPAC coefficient and hence whether attenuation coefficients could be inferred from it.
However, their analysis was limited to forward calculations of the theoretical relationships, and they did not propose an estimation procedure.

Building on the formulation of \citet{Kimura2026a}, we propose a method for estimating attenuation coefficients from observed records.
We first conduct a theoretical analysis of directional aliasing, a major obstacle to attenuation estimation, to guide the choice of array geometry.
We then present a two-stage procedure for estimating phase velocity and the attenuation coefficient using a first-order approximation of the attenuation-aware SPAC coefficient with respect to the attenuation coefficient.
Next, we apply the proposed method to ideal synthetic data and synthetic data containing various sources of error to compare the effects of these errors on attenuation and phase-velocity estimation.
We further validate the method using simulated data generated with a forward model that differs from the model assumed in the estimation procedure.
We also compare the proposed method with the Abbas method in both the verification and validation analyses.
Finally, we present a preliminary application to field data.

\section{Method}\label{s:method}

In microtremor array surveys, the vertical component of the target wavefield is assumed to consist of a superposition of mutually uncorrelated Rayleigh waves arriving from arbitrary directions with arbitrary intensities.
Such a wavefield is mathematically modelled as a three-dimensional stochastic process that is stationary in time and space, and the Rayleigh-wave phase velocity is estimated from its statistical properties.
\citet{Kimura2026a} extended this formulation to model a wavefield in which Rayleigh waves undergo exponential attenuation.
In this model, a sample path of the vertical component at position $\bm{x} = (x, y)$ is expressed as follows:
\begin{align}
  Z(t,\bm{x}) =
  \int_{-\pi}^{\pi} \int_{0}^{\infty} \int_{-\infty}^{\infty}
  \mathrm{e}^{\mathrm{i}\omega t} 
  \mathrm{e}^{- \mathrm{i}\bm{x} \cdot \bm{p} k(\omega)}
  \mathrm{e}^{-\bm{x} \cdot \bm{p} \alpha(\omega)}
  \mathrm{d}\zeta(\omega, k, \phi).
  \label{e:spectral_rep}
\end{align}
Here, $\mathrm{i}$ is the imaginary unit, $\mathrm{e}$ is Napier's constant, $\omega$ is the angular frequency, $k(\omega)$ is the wavenumber, $\alpha(\omega)$ is the attenuation coefficient, $\bm{p}$ is the unit vector representing the wave arrival direction given by $\bm{p} = (\cos\phi, \sin\phi)$ and $d\zeta$ is an increment representing the amplitude and phase of each component wave.
The local increment $\mathrm{d}\zeta(\bm{x};\omega,k,\phi)$ denotes the product of $\mathrm{d}\zeta(\omega,k,\phi)$ and the two spatial factors in Eq.~\eqref{e:spectral_rep}.
The following frequency-domain expressions concern $\omega>0$; the negative-frequency part of the real-valued wavefield is understood through conjugate symmetry.

The quantity corresponding to the spatial autocovariance function for a spatial lag $\bm{r} = (r\cos \theta,\allowbreak r\sin\theta)$ from the reference position $\bm{x}_0 = (x_0, y_0)$ is then defined as follows:
\begin{align}
  G(\bm{x}_0; \omega, r, \theta) 
  &=
    \mathcal{F}\left[E \left[ 
    Z^{*}(0,\bm{x}_0 + \bm{r}) Z(\tau, \bm{x}_0)
    \right]\right]
  \notag
  \\
  &=
  \int_{-\pi}^{\pi}
  \mathrm{e}^{\mathrm{i}k(\omega)r \cos (\phi - \theta)}
  \mathrm{e}^{- \alpha (\omega) r \cos (\phi - \theta)}
  f(\bm{x}_0; \omega, \phi) 
  \mathrm{d}\phi,
  \label{e:autocov_fn}
\end{align}
where $\mathcal{F}$ denotes the Fourier transform with respect to time lag $\tau$, with kernel $\mathrm{e}^{-\mathrm{i}\omega\tau}$, and $E$ denotes the ensemble average.
Here, the frequency-wavenumber-direction spectral density (FWDSD) $F$ of the local increments is concentrated at $k=k(\omega)$:
\begin{align}
F(\bm{x}_0; \omega, k, \phi)
  = 
  f(\bm{x}_0; \omega, \phi) \frac{\delta (k - k(\omega))}{k},
  \label{e:fd_spec}
\end{align}
where $f(\bm{x}_0; \omega, \phi)$ is the frequency-direction spectral density (FDSD)~\cite{Henstridge1979}.

The normalised frequency-direction spectral density (NFDSD) is defined by normalising the FDSD:
\begin{align}
  \lambda(\bm{x}_0; \omega, \phi) 
  = 
  \frac{
    f(\bm{x}_0; \omega, \phi)  
  }{
    \int_{-\pi}^{\pi} f(\bm{x}_0; \omega, \phi') \mathrm{d} \phi'
  }.
  \label{e:nfdsd}
\end{align}
At frequencies with nonzero power, $\int_{-\pi}^{\pi}\lambda(\bm{x}_0;\omega,\phi)\,\mathrm{d}\phi=1$.

Using the NFDSD, the quantity corresponding to the autocorrelation function is expressed as
\begin{align}
  \gamma(\bm{x}_0; \omega, r, \theta) 
  =
  \frac{G(\bm{x}_0;\omega,r,\theta)}{G(\bm{x}_0;\omega,0,\theta)}
  =
  \int_{-\pi}^{\pi}
  \mathrm{e}^{\mathrm{i}k(\omega)r \cos (\phi - \theta)}
  \mathrm{e}^{- \alpha (\omega) r \cos (\phi - \theta)}
  \lambda(\bm{x}_0; \omega, \phi) 
  \mathrm{d}\phi,
  \label{e:autocor_fn}
\end{align}

Furthermore, integrating $\gamma$ over azimuth $\theta$ yields the attenuation-aware SPAC coefficient:
\begin{align}
  \rho (\omega, r) 
  = 
  \frac{1}{2\pi}\int_{-\pi}^{\pi}
  \gamma(\bm{x}_0; \omega, r, \theta) \mathrm{d}\theta
  =
  J_0((k(\omega) + \mathrm{i}\alpha(\omega))r).
  \label{e:spac}
\end{align}

Next, we show how the SPAC coefficient is obtained using an array consisting of a finite number of observation points.
Following the supporting information of \citet{Kimura2026a}, we consider a series expansion of the spatial autocorrelation function $\gamma$.

We first expand the NFDSD in a Fourier series with respect to azimuth $\phi$ and define the Fourier coefficients $\Lambda_m$:
\begin{align}
  \Lambda_m (\bm{x}_0; \omega)
  = 
  \int_{-\pi}^{\pi}
  \mathrm{e}^{-\mathrm{i}m\phi} \lambda(\bm{x}_0; \omega, \phi) \mathrm{d}\phi.
  \label{e:Fourier_NFDSD}
\end{align}

With $X_m = \Re \Lambda_m$ and $Y_m = \Im \Lambda_m$, the spatial autocorrelation function is expanded in a series using Bessel functions of the first kind as follows.
\begin{align}
  \gamma(\bm{x}_0; \omega, r, \theta)
  =& 
  J_0((k+\mathrm{i}\alpha)r) 
  \notag\\
  &+ 
  2 \sum_{m = 1}^{\infty} 
  \mathrm{i}^m J_m ((k+ \mathrm{i}\alpha)r)
  (X_m \cos m\theta - Y_m \sin m\theta)
  \label{e:Bessel_series}
\end{align}

Next, we consider approximating the SPAC coefficient using observations from a discrete circular array, specifically a regular $N$-sided polygonal array with $\theta = \theta_0 + \frac{2\pi}{N}n\;(n=0,1,2,\ldots, N-1)$.
Following the convention of the SPAC method, $N$ is taken to be odd.
Replacing the azimuthal integral in Eq.~\eqref{e:spac} by the average over the $N$ points and using Eq.~\eqref{e:Bessel_series} gives
\begin{align}
  \hat{\rho}(\bm{x}_0; \omega, r)
  =& 
  J_0((k+\mathrm{i}\alpha)r) 
  + 
  \frac{2}{N} \sum_{m = 1}^{\infty}
  \mathrm{i}^m J_m ((k+ \mathrm{i}\alpha)r)
  \notag\\
  &\sum_{n=0}^{N-1} \left(
  X_m \cos \left(m\left(\theta_0 +\frac{2\pi}{N}n\right)\right)
  - 
  Y_m \sin \left(m\left(\theta_0 +\frac{2\pi}{N}n\right)\right)
  \right)
  \notag
  \\
  =&
  J_0((k+\mathrm{i}\alpha)r) 
  + 
  \notag\\
  &2 \sum_{l = 1}^{\infty} 
  \mathrm{i}^{lN} J_{lN} ((k+ \mathrm{i}\alpha)r)
  \left(X_{lN} \cos(lN\theta_0) - Y_{lN} \sin(lN\theta_0) \right)
  \label{e:d_alias}
\end{align}

In Eq.~\eqref{e:d_alias}, the summation following the first term represents the error relative to the theoretical value $J_0((k + \mathrm{i}\alpha)r)$ and is referred to as directional aliasing because $X_{lN}=Y_{lN}=0$ in an isotropic wavefield.

With $A_m = X_m\cos(m\theta_0) - Y_m\sin(m\theta_0)$, Eq.~\eqref{e:d_alias} can be rewritten as follows:
\begin{align}
\hat{\rho}(\bm{x}_0;\omega,r)
={}&
J_0\left((k+\mathrm{i}\alpha)r\right)
\notag\\
&+
2\sum_{l=1}^{\infty}
(-1)^l
J_{2lN}\left((k+\mathrm{i}\alpha)r\right)
A_{2lN}
\notag\\
&+
2\mathrm{i}(-1)^{(N-1)/2}
\sum_{l=1}^{\infty}
(-1)^{l-1}
J_{(2l-1)N}\left((k+\mathrm{i}\alpha)r\right)
A_{(2l-1)N}.
\end{align}

\begin{figure*}[tbp]
\centering
  \includegraphics[width=\textwidth]{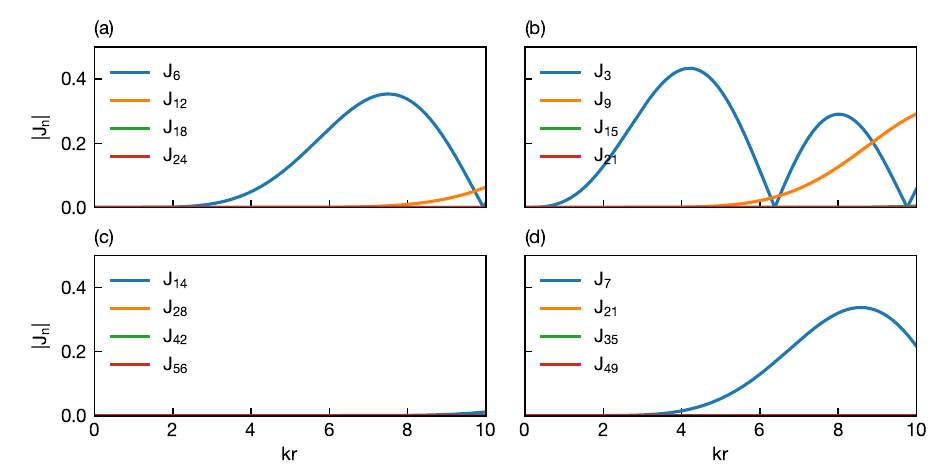}
\caption{Magnitudes of the Bessel-function factors in the azimuthal-aliasing terms for regular polygonal arrays in the zero-attenuation limit. (a, b) Triangular array ($N=3$). (c, d) Heptagonal array ($N=7$). The left and right columns show the factors contributing to the real and imaginary parts of the SPAC coefficient, respectively. The first four orders are plotted in each panel: $2lN$ for the real part and $(2l-1)N$ for the imaginary part, with $l=1,\ldots,4$.}\label{fig01}
\end{figure*}

The conventional SPAC method assumes a wavefield without attenuation, i.e. $\alpha=0$, so the real part is affected only by directional aliasing from the terms of order $2lN$.
Assuming that attenuation effects are sufficiently small, a zeroth-order approximation gives $J_m\left((k+\mathrm{i}\alpha)r\right)\simeq J_m\left(kr\right) \in \mathbb{R}$, showing that the real and imaginary parts of the aliasing terms contain Bessel functions of order $2lN$ and $(2l-1)N$, respectively.
Since $|\Lambda_m| \leq 1$~\cite{Kimura2023b}, $|A_m| \leq 1$, and the effect of directional aliasing is largely determined by the values of $J_{lN}$.
For $n>0$ and $x\in\mathbb{R}$, $J_{n}(x)$ is known to satisfy $J_{n}(x) \simeq 0$ in the range $x \ll n$ (see Fig.~\ref{fig01}).
Thus, directional aliasing can be neglected where $kr$ is sufficiently small in relative terms.
When phase velocity is estimated using the SPAC method with a conventional triangular array, $N=3$ and the first aliasing term in the real part is $J_6$, so the range corresponding to the first trough of the $J_0$ function, $kr \simeq 3.8 < 6$, is conventionally assumed to be aliasing free.

However, \citet{Kimura2026a} pointed out that attenuation effects appear more strongly in the imaginary part than in the real part, suggesting that attenuation estimation is difficult with a triangular array, for which the first aliasing term is $J_3$.
In this paper, we therefore use a regular heptagonal array for attenuation estimation.
Its first aliasing term has an effect no greater than that of $J_6$, which is considered acceptable for the real part in the conventional SPAC method.
Fig.~\ref{fig01} shows the absolute values of the Bessel-function factors in the real and imaginary parts of the aliasing terms for regular triangular and heptagonal arrays, plotted against $kr$ under the zeroth-order approximation $\alpha \simeq 0$.
A comparison of Figs~\ref{fig01}(a) and (d) shows that the $J_n$ functions governing the aliasing effects are broadly comparable for the real part with a triangular array and the imaginary part with a regular heptagonal array, with slightly smaller absolute values for the latter.

Based on the above considerations, this paper proposes a procedure for estimating the attenuation coefficient $\alpha(\omega)$ and phase velocity $c(\omega) = \omega / k(\omega)$ from the SPAC coefficient $\tilde{\rho}(\omega,r)$ calculated from observed records.
Specifically, the relationship in Eq.~\eqref{e:spac} is used without explicitly accounting for directional aliasing arising from the finite number of observation points.
Although $c$ and $\alpha$ could be estimated directly by fitting $\tilde{\rho}(\omega,r)$, we use an estimation procedure that takes advantage of the tendency for phase-velocity effects to appear in the real part and attenuation effects in the imaginary part.
For $\alpha r \ll 1$, a first-order approximation of Eq.~\eqref{e:spac} gives
\begin{align}
  \rho(\omega, r) \simeq J_0(kr) - \mathrm{i}\alpha r J_1 (kr).
\end{align}
This gives $\Re \rho \simeq J_0(kr)$, the same relationship as in the SPAC method without attenuation.
Using this relationship, phase velocity and attenuation coefficient are obtained through the following procedure:
\begin{enumerate}
  \item Calculate $\tilde{\rho}(\omega, r)$ from the observed records.
  \item Apply the conventional SPAC method to $\Re \tilde{\rho}(\omega, r)$ and use the resulting phase velocity as the initial value $c_{\mathrm{init}}$.
  \item Use $\alpha_{\mathrm{init}} = -\Im \tilde{\rho}(\omega, r) / \left[r J_1 (\omega r / c_{\mathrm{init}})\right]$ as the initial value of the attenuation coefficient where the denominator is nonzero.
  \item Using the initial values $c_{\mathrm{init}}$ and $\alpha_{\mathrm{init}}$, apply an optimisation method to obtain $c$ and $\alpha$ that minimise $\mathrm{misfit} = \left| \tilde{\rho}(\omega, r) - J_0\left(\left(\frac{\omega}{c} + \mathrm{i}\alpha\right) r\right)\right|^2$.
\end{enumerate}

Differential evolution~\cite{Storn1997} was adopted as the optimisation method because it is effective for optimising multimodal functions and relatively robust against disturbances such as noise.
The pair of values $c_{\mathrm{init}}$ and $\alpha_{\mathrm{init}}$ was used as the initial best solution.

\section{Verification using synthetic data}\label{s:ver}

The basic functionality of the proposed method was first verified using synthetic waveforms generated with the model in Eq.~\eqref{e:spectral_rep} as the forward model.
This verification is insufficient to establish the validity of the proposed method because the same model is used to synthesise the waveforms and estimate phase velocity and attenuation coefficient.
Nevertheless, it is useful to know how well the proposed method performs when the actual microtremor field is known to follow the model, and it is also necessary to examine whether inversion is mathematically feasible for realistic values of $c$ and $\alpha$.

In this section, we also conducted experiments in which we controlled the length of the records used for analysis, wavefield anisotropy and incoherent noise that may be introduced during observations.
Methods that analyse microtremors, such as the SPAC method, assume ergodicity and divide the records into short segments, treating each segment as a sample to estimate statistical quantities such as the autocorrelation function.
Thus, controlling the length of the observed records amounts to controlling the number of segments used to estimate the statistical quantities and, consequently, the stochastic error.
Controlling wavefield anisotropy also controls the directional aliasing examined in the previous section.
The controlled incoherent noise represents differences in the levels of instrumental self-noise and non-plane-wave components from sources in the immediate vicinity of observation points~\cite{Cho2021a}.

\begin{table*}[t]
\centering
\caption{Velocity, density and attenuation parameters of the two-layer model used for synthetic-data verification. The $Q$ values correspond to Cases A, B and C in that order.}\label{tab01}
\begin{tabular}{ccccc}
  Thickness (m) & $V_\mathrm{P}$ ($\mathrm{m\,s^{-1}}$) & $V_\mathrm{S}$ ($\mathrm{m\,s^{-1}}$) & Density ($\mathrm{kg\,m^{-3}}$) & $Q_\mathrm{P} = Q_\mathrm{S}$ (Cases A, B and C)\\\hline
  20 & 300 & 150 & 1700 & 15, 30 and 60\\
  $\infty$ & 2000 & 1000 & 2000 & 45, 90 and 180\\
\end{tabular}
\end{table*}

\begin{figure*}[tbp]
\centering
\includegraphics[width=\textwidth]{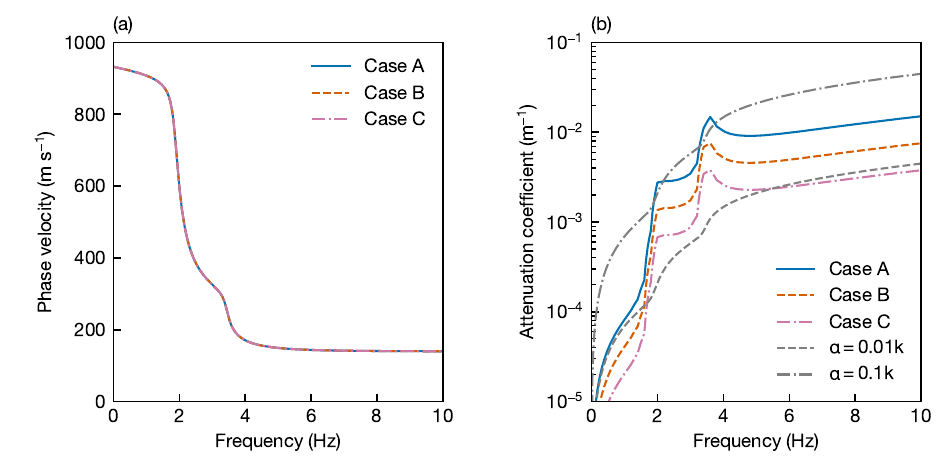}
\caption{Theoretical fundamental-mode Rayleigh-wave (a) phase velocities and (b) attenuation coefficients for three attenuation levels with a common velocity and density structure. As labelled in the figure, Cases A, B and C correspond to high, intermediate and low attenuation, respectively. The grey curves in (b) indicate $\alpha=0.01k$ and $\alpha=0.1k$, where $k=\omega/c$.}\label{fig02}
\end{figure*}

To generate the synthetic data, we assumed the simple two-layer structure shown in Table~\ref{tab01}.
The velocity and density structure was common to all cases, and three attenuation cases, A, B and C in order of decreasing attenuation, were prepared.
In each case, $Q_{\mathrm{S}}=Q_{\mathrm{P}}$, and the Q value of the lower layer was set to three times that of the upper layer.
Fig.~\ref{fig02} shows the phase-velocity dispersion curves and attenuation coefficient curves calculated from this velocity, density and attenuation structure using an attenuation-aware generalised R/T method (gRTm) that combines the methods of Ref.~\cite{Pan2022} and~\cite{Pei2008,Pei2009}.
Fig.~\ref{fig02}(b) also shows curves for $\alpha=0.01k$ and $\alpha=0.1k$, where $k=\omega/c$ is the wavenumber.
These values correspond to the low and high attenuation levels considered by \citet{Nakahara2012}.
The attenuation values used in this study broadly span this range but tend towards its lower end, providing relatively demanding verification conditions.

The synthetic data were generated following the method of \citet{Kimura2023b}, with the NFDSD controlled as a discrete distribution of arrival directions.
Note that propagation in space was treated deterministically rather than stochastically.
The von Mises distribution~\cite{Mardia1999} was used as the functional form of the NFDSD, with the direction parameter $\mu$ set to $\pi/6$ in all cases.
The actual arrival directions were specified by 1000 realisations drawn from the von Mises distribution, and time histories of 1,025,024 samples were generated at 50\,sps.
These time histories were divided into 1000 segments of 40.96\,s with 50\% overlap.
Each segment was windowed using a half-cycle cosine window, and statistical quantities such as the autocovariance and autocorrelation functions were calculated by ensemble averaging.
In each case, records were synthesised at the seven vertices and centre of a regular heptagonal array, comprising 8 observation points in total.
Two array sizes, with radii of 30.0\,m and 58.5\,m, were used.
The initial phase velocity was determined by dividing the $J_0$ function into monotonic intervals, applying its inverse to $\Re\tilde{\rho}(\omega, r)$ to obtain possible phase-velocity candidates~\cite{Kimura2026b} and selecting the candidate closest to the theoretical dispersion curve at each frequency.
Thus, the phase-velocity estimates were not obtained solely from the observed records.
This procedure was used to automatically process the results for many cases, and similar results can readily be obtained in practice using a procedure such as that in Section~\ref{s:pre}.

\subsection{Verification of basic performance}\label{ss:basics}

First, to examine the basic performance, numerical experiments were conducted with the concentration parameter set to $\kappa=0$ for an ``isotropic case'' and $\kappa = 11.36$ for an ``anisotropic case''.
The former is an isotropic wavefield with a theoretical mean resultant length of $\mathrm{MRL}=0$, whereas the latter is a nearly unidirectional wavefield with $\mathrm{MRL} \fallingdotseq 0.955$, close to 1.

\begin{figure*}[tbp]
\centering
\includegraphics[width=\textwidth]{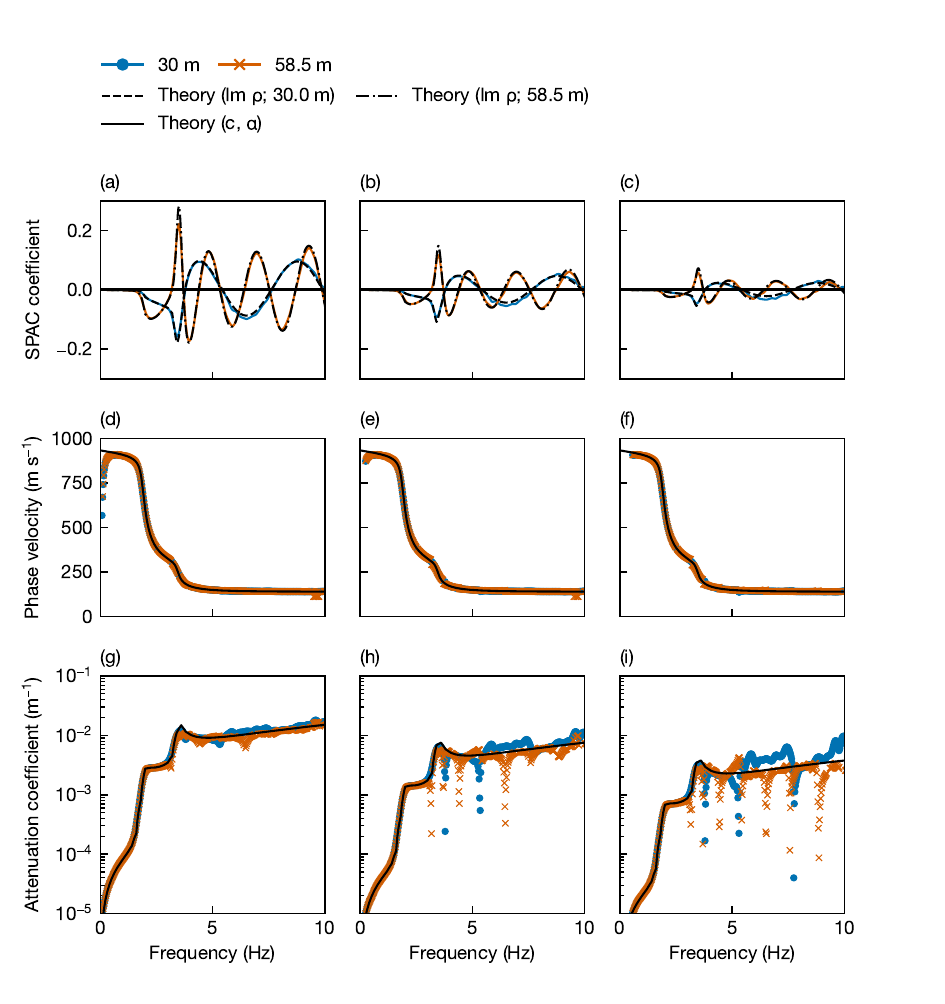}
\caption{Verification for isotropic wavefields without added noise, using 1000 segments. Columns show Cases A, B and C (high to low attenuation). (a--c) Imaginary parts of SPAC coefficients. (d--f) Phase velocities. (g--i) Attenuation coefficients. Blue and orange denote array radii of 30 and 58.5\,m; black curves show theoretical values. Differential evolution estimates are retained where each SPAC component is reproduced within 1\% relative error.}\label{fig03}
\end{figure*}

Fig.~\ref{fig03} shows the imaginary parts of the SPAC coefficients, phase velocities and attenuation coefficient curves estimated from synthetic waveforms for the three cases with different attenuation levels.
In these cases, the wavefield was isotropic, no incoherent noise was added and all 1000 segments were used to calculate the statistical quantities.
These are therefore ideal cases with negligible directional aliasing, incoherence and stochastic error.
Figs~\ref{fig03}(a)--(c) show that the imaginary parts of the SPAC coefficients are almost identical to the theoretical values.
Some deviations from the theoretical values are observed, particularly at higher frequencies, which are considered to arise from slight directional aliasing because only a finite number of arrival directions were sampled, so the resulting wavefield was not perfectly isotropic.
Figs~\ref{fig03}(d)--(f) show that phase velocities are correctly estimated at almost all frequencies.
Figs~\ref{fig03}(g)--(i) show that the overall trends of the attenuation coefficients are generally estimated correctly.
In particular, the estimates agree closely with the theoretical values at lower frequencies.

However, approximately periodic drops in the estimates are observed at higher frequencies, particularly when the attenuation level is low.
Systematic overestimation is also observed at higher frequencies in Case C.
Both phenomena are considered to result from the low sensitivity of the SPAC coefficient to the attenuation coefficient, $\partial \rho / \partial \alpha = -\mathrm{i}r J_1((k+\mathrm{i}\alpha)r)$.
First, the sensitivity exhibits approximately periodic variations owing to the oscillatory behaviour of the $J_1$ function, and the drops in the estimates are considered to occur at the sensitivity minima.
Furthermore, because the sensitivity is proportional to the array size $r$, a smaller array has lower sensitivity and is considered to be more susceptible to slight directional aliasing, which appears to cause the deviations of the estimates from the theoretical values at higher frequencies for the 30\,m array.
The relationship between sensitivity and the estimates is discussed in detail in Section~\ref{ss:mrl_sweep}.
These results show that the proposed method generally functions as expected under ideal conditions without noise or other disturbances, although problems with the estimates may arise at higher frequencies.

\begin{figure*}[tbp]
\centering
\includegraphics[width=\textwidth]{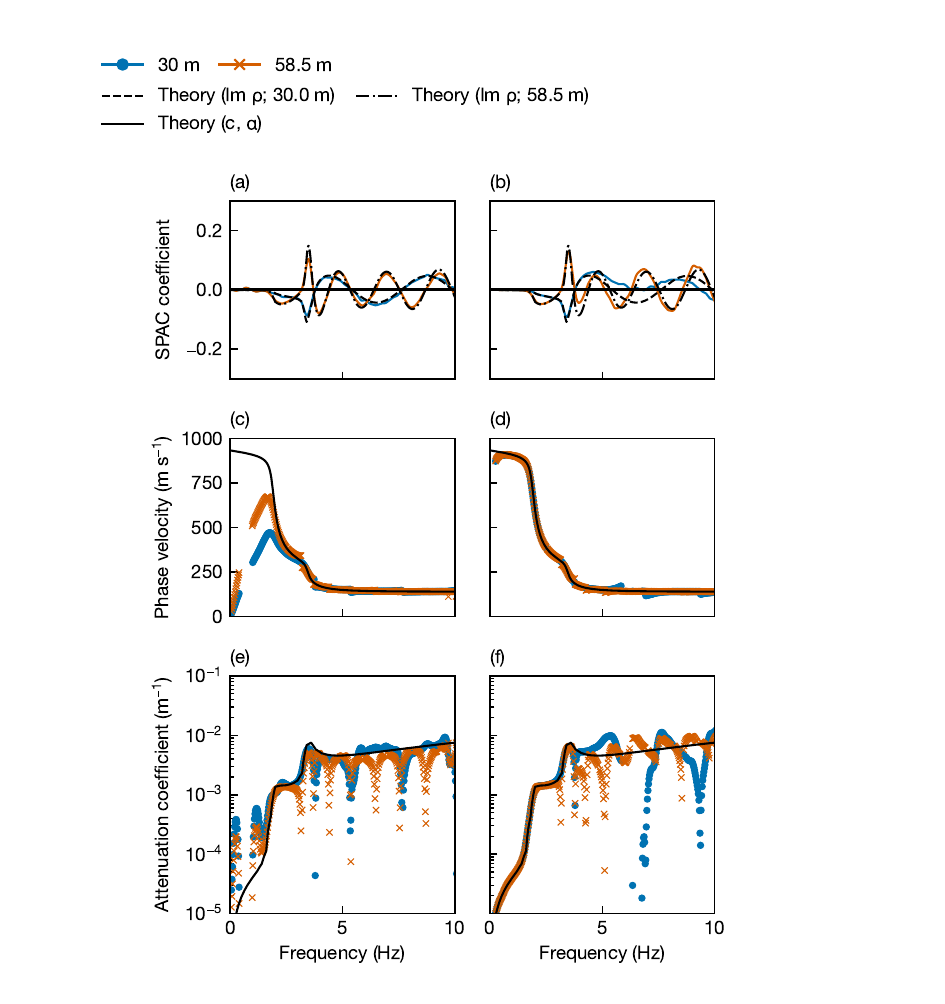}
\caption{Effects of noise and anisotropy for Case B with 1000 segments. Left: isotropic wavefield with independent Gaussian white noise (standard deviation: 30\% of signal RMS). Right: anisotropic wavefield ($\kappa=11.36$, theoretical MRL $\approx0.955$) without added noise. Rows show imaginary parts of SPAC coefficients, phase velocities and attenuation coefficients. Colours, array radii, theoretical curves and the 1\% SPAC fit criterion follow Fig.~\ref{fig03}.}\label{fig04}
\end{figure*}

Fig.~\ref{fig04} shows the results for Case B with either added incoherent noise or an anisotropic wavefield.
For the incoherent noise, Gaussian white noise with a standard deviation of 30\% of the waveform RMS was independently added to the records at each observation point.
Figs~\ref{fig04}(a) and (b) show that incoherent noise appears to have a relatively small effect on the imaginary part of the SPAC coefficient, whereas an anisotropic wavefield causes large deviations from the theoretical values at higher frequencies.
Fig.~\ref{fig04}(c) shows that incoherent noise contributes to the underestimation of phase velocity at lower frequencies, as is well known.
In contrast, Fig.~\ref{fig04}(d) shows that the phase-velocity estimates agree with the theoretical values at almost all frequencies, indicating that directional aliasing has a limited effect on phase-velocity estimation.
Turning to Fig.~\ref{fig04}(e), the attenuation estimates deviate considerably from the theoretical values at lower frequencies.
However, the lowest frequency at which attenuation coefficients are correctly estimated is lower than that for phase velocity.
This result provides a basis for regarding attenuation estimates as reliable at lower frequencies where phase velocities are correctly estimated.
The periodic drops in the estimates observed in Fig.~\ref{fig03}(h) also appear to be more pronounced in Fig.~\ref{fig04}(e).
This is an expected result because a reduction in the signal-to-noise ratio (SNR) due to incoherent noise and low sensitivity are assumed to have comparable effects.
Finally, Fig.~\ref{fig04}(f) shows that the estimates oscillate considerably at higher frequencies.
Interestingly, despite these large oscillations, their centres appear to be broadly consistent with the theoretical values.
Overall, incoherent noise affects the estimates at lower frequencies, with a greater effect on phase velocity than on attenuation coefficient.
In contrast, directional aliasing affects attenuation coefficient estimates at higher frequencies but has almost no effect on phase-velocity estimates.

\begin{figure*}[tbp]
\centering
\includegraphics[width=\textwidth]{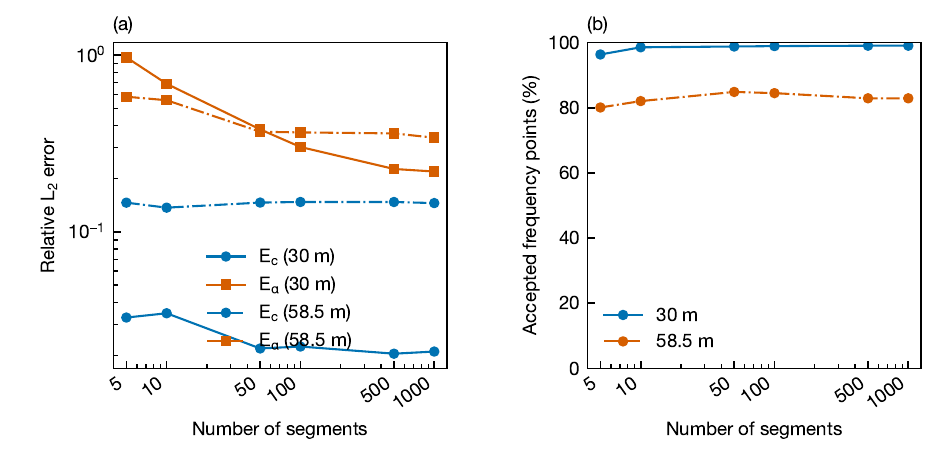}
\caption{Effect of the number of segments on estimates for isotropic Case B without added incoherent noise. (a) Relative $L_2$ errors of phase velocity ($E_c$) and attenuation coefficient ($E_\alpha$), evaluated only at accepted frequencies. Solid and dash-dotted lines represent the 30 and 58.5\,m arrays, respectively. (b) Percentage of all analysed frequency points accepted for each array. A frequency point is accepted when the reconstructed real and imaginary SPAC components each have a relative error of 1\% or less with respect to the corresponding component calculated from the records.}\label{fig05}
\end{figure*}

To evaluate the effect of stochastic error, we examined how the accuracy of phase-velocity and attenuation coefficient estimates changes with the number of segments used to estimate the statistical quantities.
Fig.~\ref{fig05}(a) plots the relative L2 errors against the number of segments.
For phase velocity, the improvement in accuracy with an increasing number of segments is limited for both array sizes.
The marginal benefit appears to decrease beyond approximately 50 segments.
In contrast, the L2 errors for the attenuation coefficient continue to decrease as the number of segments increases from approximately 50 to 1000, suggesting that increasing the number of segments, i.e. extending the observation duration, is more effective for attenuation estimation than for phase-velocity estimation.
Fig.~\ref{fig05}(b) shows the proportion of all frequency points at which the real and imaginary parts of the SPAC coefficient reconstructed from $(c, \alpha)$ estimated by the proposed method fit the corresponding components calculated from the observed waveforms.
Specifically, this is the proportion of frequency points at which the reconstruction errors of both the real and imaginary parts are within 1\%, i.e. $\left|\Re\rho_{\mathrm{rec}} - \Re\rho_{\mathrm{obs}} \right| \leq 0.01 |\Re\rho_{\mathrm{obs}}|$, with the same condition applied to the imaginary part.
The relative L2 errors in Fig.~\ref{fig05}(a) were calculated using only the frequency points satisfying this condition.
The proportion of accepted frequency points in Fig.~\ref{fig05}(b) remains almost constant as the number of segments increases.
The decrease in the relative L2 errors in Fig.~\ref{fig05}(a) therefore suggests improved accuracy among the accepted estimates rather than an increase in the number of frequencies at which acceptable solutions are obtained.
This may reflect reduced stochastic perturbations of the local misfit landscape as the number of segments increases, leading to more accurate parameter estimates.
Thus, the effects of obvious outliers are not included in Fig.~\ref{fig05}(a).
These results suggest that attenuation coefficient estimation requires longer observation durations than are normally used for phase-velocity estimation.

\begin{figure*}[tbp]
\centering
\includegraphics[width=\textwidth]{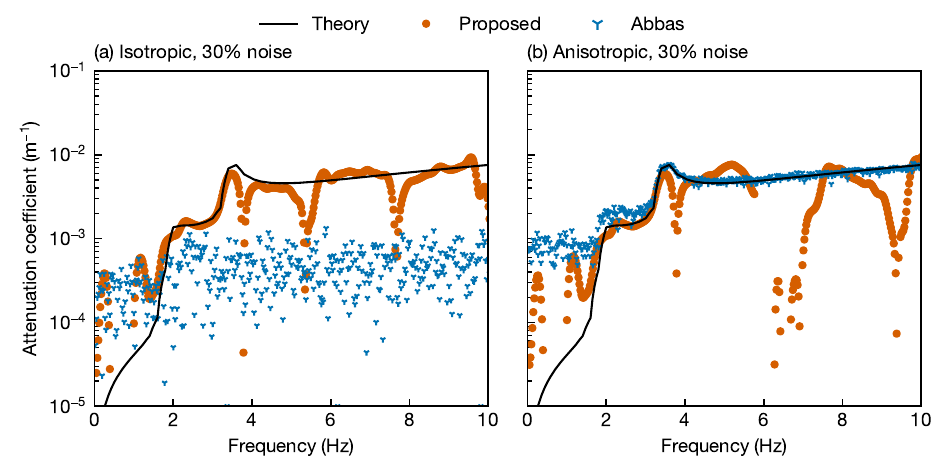}
\caption{Comparison of attenuation coefficients estimated by the proposed and Abbas methods for Case B using a 30\,m-radius array and 1000 segments. (a) Isotropic wavefield. (b) Anisotropic wavefield ($\kappa=11.36$, theoretical MRL approximately 0.955). Both cases include independent Gaussian white noise with a standard deviation of 30\% of the signal RMS. Black curves show theoretical attenuation coefficients. Proposed-method estimates satisfy the component-wise 1\% SPAC fit criterion.}\label{fig06}
\end{figure*}

Finally, we compare the proposed and Abbas methods for the isotropic and anisotropic cases of Case B.
To provide a more realistic setup, 30\% incoherent noise was added in both cases.
Fig.~\ref{fig06}(a) shows that, for an isotropic wavefield, the proposed method successfully estimates values close to the theoretical values except at very low frequencies, whereas the Abbas method estimates a constant value that deviates considerably from the theoretical values across the entire frequency range.
Because the Abbas method theoretically assumes a unidirectional wavefield, its limited applicability to an isotropic wavefield is expected.
Fig.~\ref{fig06}(b) shows that, for an anisotropic wavefield, the Abbas method provides highly stable estimates that agree with the theoretical values, whereas the proposed method gives unstable estimates at higher frequencies.
As discussed above, the instability of the proposed method is caused by directional aliasing, whereas the Abbas method is unaffected, possibly because it assumes a unidirectional wavefield.
However, the proposed method also provides stable estimates sufficiently close to the theoretical values at approximately 2--4\,Hz and is more accurate than the Abbas method at 2--3\,Hz.
This suggests that the wavelength ranges over which stable estimates can be obtained may differ for the same array size.
These results suggest that the proposed and Abbas methods perform well over different wavelength ranges and for different degrees of concentration of wave arrival directions.
Because the two methods differ only in their analysis procedures and can be applied to the same array records, their combined use may be effective for estimating attenuation coefficients.

\subsection{Detailed examination of wavefield anisotropy}\label{ss:mrl_sweep}

\begin{figure*}[tbp]
\centering
\includegraphics[width=\textwidth]{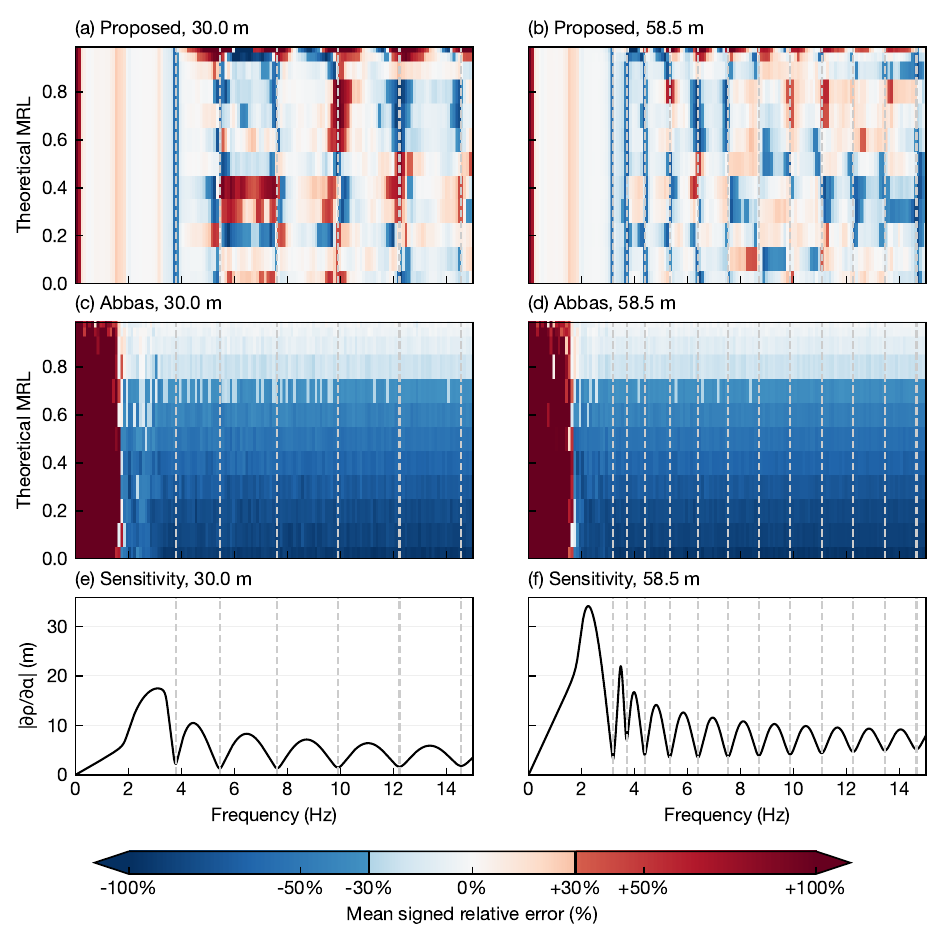}
\caption{Dependence of attenuation estimates on theoretical MRL and array radius for Case B without added incoherent noise, using 1000 segments. (a, b) Proposed method. (c, d) Abbas method using BFM. Colours indicate the signed relative error $(\hat{\alpha}-\alpha)/\alpha$, averaged over 0.1\,Hz bins; pale colours indicate errors within $\pm30\%$. (e, f) Magnitude of the SPAC sensitivity to attenuation, $|\partial\rho/\partial\alpha|=r|J_1((k+\mathrm{i}\alpha)r)|$, evaluated using theoretical values. Left and right columns correspond to radii of 30 and 58.5\,m, respectively. Vertical dashed lines mark local sensitivity minima.}\label{fig07}
\end{figure*}

To examine the effects of wavefield anisotropy in more detail, synthetic data were generated using the same subsurface model as in Case B, with theoretical MRL values of $\mathrm{MRL} = 0, 0.1, 0.2,\ldots, 0.9, 0.95, 0.975$.
Fig.~\ref{fig07} shows the signed relative errors of the estimates from the proposed and Abbas methods with respect to the theoretical values, averaged over 0.1\,Hz bins, for the two array sizes used in the previous subsection.

First, Figs~\ref{fig07}(a) and (b) show that the proposed method has broadly similar estimation accuracy across all MRL values, with particularly good agreement between the estimates and theoretical values at frequencies of 4 Hz or lower.
The estimates from the proposed method oscillate at higher frequencies.
In this higher-frequency range, the estimates alternate between underestimation and overestimation and tend to agree with the theoretical values in between.
Figs~\ref{fig07}(c) and (d) show that the estimates from the Abbas method agree well with the theoretical values only at high MRL values and are stable at higher frequencies but deviate at frequencies of 2\,Hz or lower.
These results further support the findings suggested in Section~\ref{ss:basics}.

Figs~\ref{fig07}(e) and (f) show the sensitivity of the SPAC coefficient to the attenuation coefficient.
The sensitivity is generally greater for the larger array and oscillates with frequency.
Vertical dashed lines in Figs~\ref{fig07}(a)--(d) mark the sensitivity minima, and the errors in the estimates from the proposed method are larger around these frequencies.
This further confirms that the scatter in the estimates at higher frequencies is associated with local reductions in sensitivity.
The frequencies at which the sensitivity minima occur also differ with array size.
Thus, combining estimates from multiple array sizes may provide more accurate estimates.

\subsection{Caution regarding the use of relative-coordinate arrays}

\begin{figure*}[tbp]
\centering
\includegraphics[width=\textwidth]{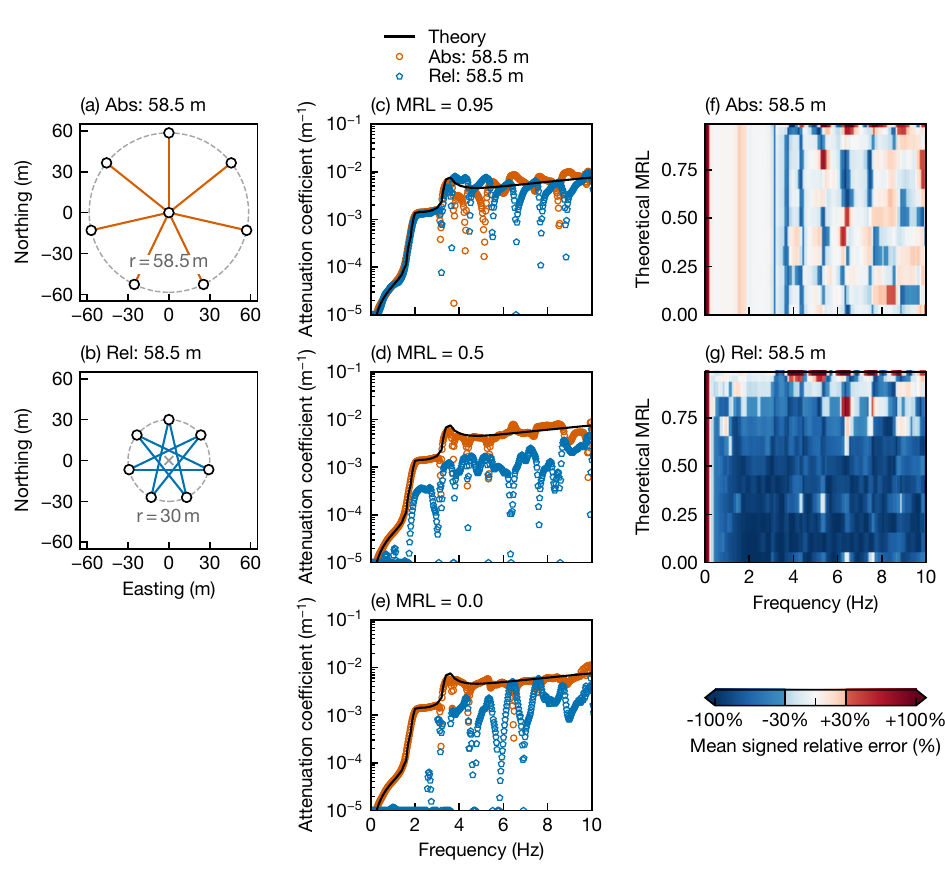}
\caption{Comparison of absolute- and relative-coordinate arrays using the proposed method. (a) Absolute-coordinate array with a radius of 58.5\,m. (b) Relative-coordinate array with a sensor-pair distance of 58.5\,m obtained from a 30\,m-radius array. (c--e) Attenuation estimates for theoretical MRLs of 0.95, 0.5 and 0, respectively. (f, g) Signed relative errors $(\hat{\alpha}-\alpha)/\alpha$ for the absolute- and relative-coordinate arrays, averaged over 0.1\,Hz bins; pale colours indicate errors within $\pm30\%$. Results use 1000 segments without added incoherent noise.}\label{fig08}
\end{figure*}

In the conventional SPAC method without attenuation, an absolute-coordinate array (ACA, Fig.~\ref{fig08}a) and a relative-coordinate array (RCA, Fig.~\ref{fig08}b) are equivalent owing to spatial stationarity.
However, when attenuation is considered, spatial stationarity no longer holds, making it problematic to treat an RCA as equivalent to an ACA.
In this subsection, we therefore use the data from Section~\ref{ss:mrl_sweep} to examine the conditions under which an RCA can and cannot be used.

Figs~\ref{fig08}(c)--(e) show the attenuation coefficients estimated using the ACA and RCA for theoretical MRL values of 0.95, 0.5 and 0.0, respectively.
The ACA yields attenuation coefficients broadly consistent with the theoretical values at different MRL values, whereas the RCA underestimates the coefficients over most frequencies when the MRL is 0.0 or 0.5 and provides results comparable to those of the ACA only when the MRL is 0.95.
This suggests that the use of an RCA is acceptable only for nearly unidirectional wavefields.
Figs~\ref{fig08}(f)--(g) show colour maps of the signed relative errors between the estimated and theoretical attenuation coefficients for different MRL values, averaged over 0.1 Hz bins.
These figures also show that the ACA estimates depend little on MRL, whereas the RCA estimates agree with the theoretical values only at high MRL values.
Fig.~\ref{fig08}(f) reproduces Fig.~\ref{fig07}(b) for comparison.
The dependence of the RCA results on wavefield directionality is examined analytically in \ref{app:rca_edge_cases}.
These results indicate that the use of relative-coordinate arrays with the proposed method is generally not recommended, and careful attention should be paid to the directional characteristics of the wavefield if such arrays are used.

\section{Validation using simulated data}\label{s:val}

In the verification in Section~\ref{s:ver}, a common model was used for waveform generation and parameter estimation.
In this section, we examine the validity of the proposed method using microtremor waveforms generated by independent numerical wave-propagation simulations.
The subsurface model of \citet{Abbas2025} was used without modification, whereas the microtremor generation procedure and the distribution of wave arrival directions were changed.
Specifically, pulse responses of the P--SV wavefield calculated using the spectral element method (SEM) were superposed to generate microtremor waveforms with controlled arrival-direction distributions, as in Section~\ref{s:ver}.
This allows us to examine whether the proposed method functions under conditions closer to those in actual microtremor fields, which may also contain body waves and surface-wave components other than the fundamental mode.

\begin{figure*}[tbp]
\centering
\includegraphics[width=\textwidth]{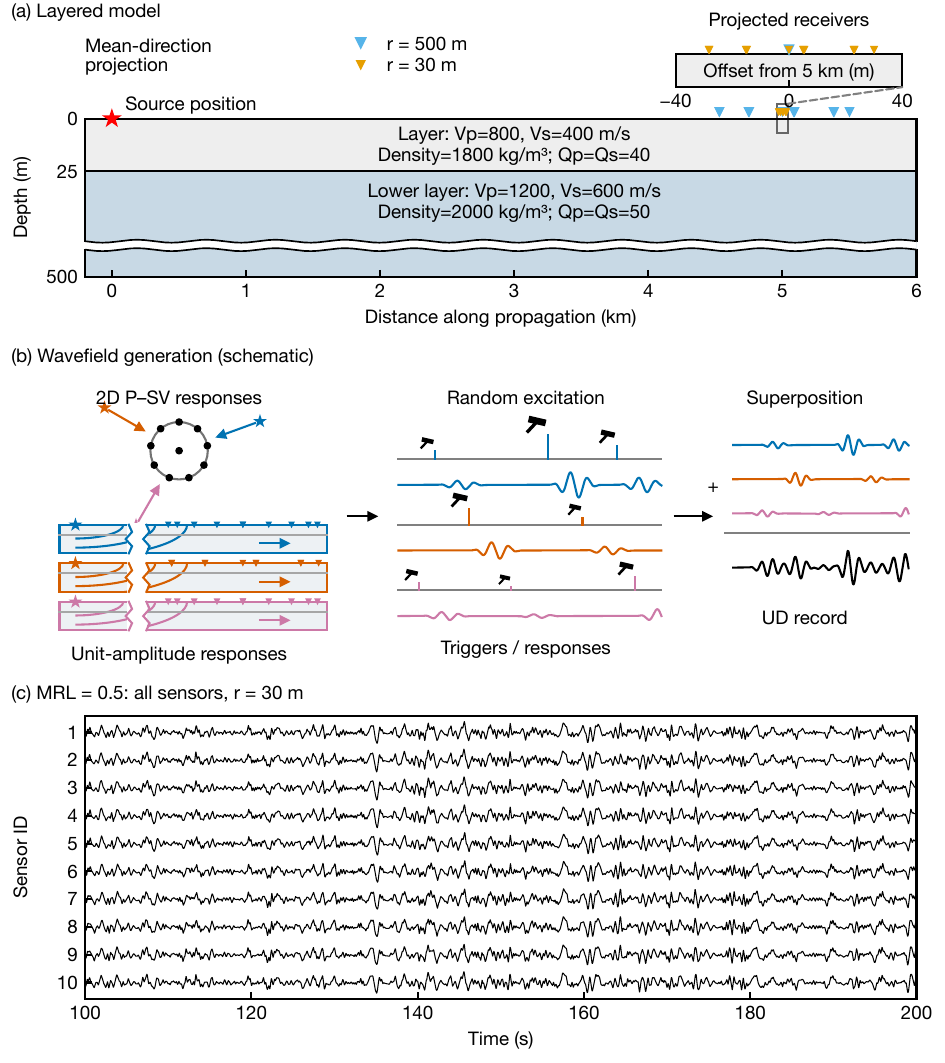}
\caption{Numerical model and generation of simulated microtremor records. (a) Layered model and projected receiver positions. (b) Superposition of P--SV pulse responses with random excitation times, amplitudes and arrival directions. (c) Example vertical-component records at the ten sensors of the 30\,m-radius array for a theoretical MRL of 0.5.}\label{fig09}
\end{figure*}

Fig.~\ref{fig09}(a) shows the subsurface model used in the simulations.
The model was essentially identical to the ``Layer above a half-space model'' used by \citet{Abbas2025} in their verification.
The velocity and attenuation structure of each layer was set as shown in Fig.~\ref{fig09}, consistent with Fig.~9 of \citet{Abbas2025}.
However, because densities were not specified in the original paper, realistic values appropriate for the velocity structure were assigned: $1800\,\mathrm{kg\,m^{-3}}$ for the surface layer and $2000\,\mathrm{kg\,m^{-3}}$ for the underlying half-space.
The numerical simulations were performed using SPECFEM2D~\cite{Komatitsch1998}, with a free surface and absorbing boundaries at the sides and bottom.
The mesh spacing was 25\,m horizontally and 5\,m and 25\,m vertically in the surface and lower layers, respectively.
Each element contained $5\times5$ Gauss--Lobatto--Legendre points, and the time step was $5.0\times10^{-4}$\,s.
The distance from the source at the ground surface to the array centre was fixed at 5\,km.

We first prepared an array with a radius of 500\,m, as used by \citet{Abbas2025}.
However, Section~\ref{ss:basics} suggested that the proposed method provides good results at longer wavelengths relative to the array size than the Abbas method.
Because a smaller array may be suitable when targeting the same frequency range, an array with a radius of 30\,m was also prepared and the two were compared.
Both arrays consisted of 10 observation points: 9 at the vertices of a regular nonagon and 1 at the centre.

The waveform generation procedure is shown in Fig.~\ref{fig09}(b).
First, vertical-component responses at the ground surface were calculated for 21 Hann bursts with a duration of 2\,s and central frequencies ranging from 0.5 to 10.5\,Hz in increments of 0.5\,Hz.
Next, the observation-point coordinates were projected onto each arrival direction.
The responses were interpolated at the corresponding distances and then superposed using random excitation times and amplitudes.
As in Section~\ref{s:ver}, the arrival directions were specified by 1000 realisations drawn from the von Mises distribution.
This generation procedure uses two-dimensional P--SV responses of a horizontally layered structure for each azimuth and does not directly simulate a three-dimensional distribution of point sources.

Fig.~\ref{fig09}(c) shows an example of the generated waveforms.
The record length was 10250.24\,s at a sampling rate of 100\,sps.
A total of 1000 segments of 20.48\,s with 50\% overlap were used, and the spectra were smoothed using a Parzen window with a bandwidth of 0.2\,Hz.

\subsection{Effects of array size and comparison with the Abbas method}\label{ss:val_basics}

\begin{figure*}[tbp]
\centering
\includegraphics[width=\textwidth]{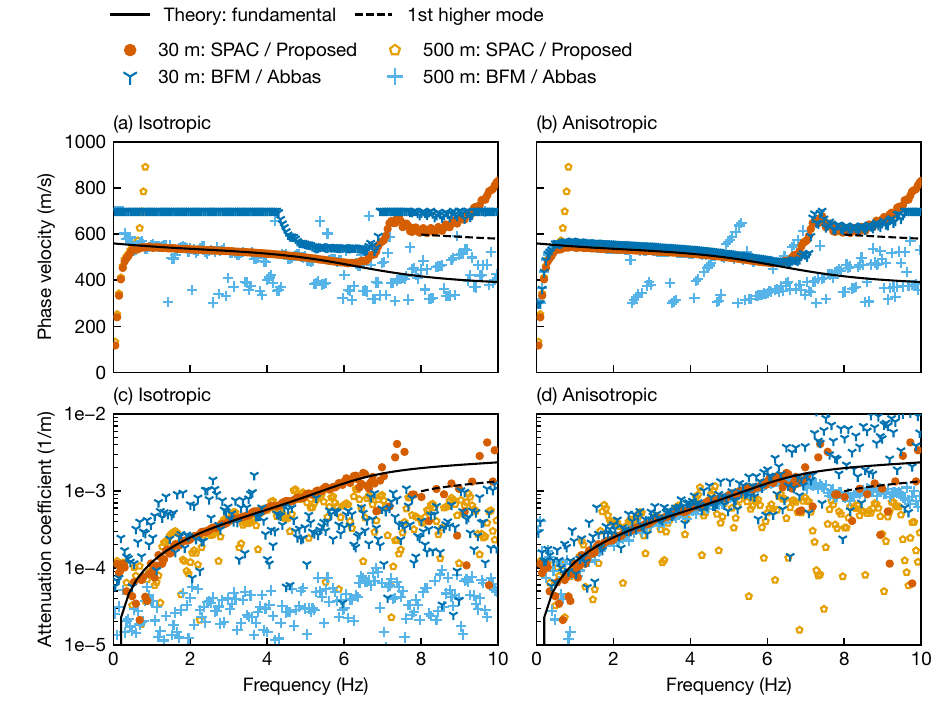}
\caption{Comparison between the 30\,m- and 500\,m-radius arrays for isotropic and anisotropic wavefields. (a, b) Phase velocities obtained by SPAC and conventional beamforming (BFM). (c, d) Attenuation coefficients estimated by the proposed and Abbas methods. Solid and dashed black curves indicate the theoretical fundamental and first higher modes, respectively. The theoretical MRL is 0 for the isotropic case and approximately 0.955 for the anisotropic case.}\label{fig10}
\end{figure*}

First, as in Section~\ref{ss:basics}, we compared an ``isotropic case'' with an ``anisotropic case'' having $\kappa=11.36$ and a theoretical MRL of approximately 0.955.
Fig.~\ref{fig10} shows the phase velocities obtained by the SPAC method and conventional beamforming (BFM) and the attenuation coefficients obtained by the proposed and Abbas methods.
The theoretical curves of the fundamental and first higher modes calculated using gRTm were used as references.

In the isotropic case in Fig.~\ref{fig10}(c), the estimates from the proposed method agree well with the theoretical fundamental-mode attenuation coefficients over approximately 1--6\,Hz, whereas the estimates from the Abbas method deviate considerably from the theoretical values at almost all frequencies.
Regarding the difference between array sizes, the proposed method gives more stable estimates with the 30\,m array than with the 500\,m array.
This is consistent with the results obtained in Section~\ref{ss:basics}.

In the anisotropic case in Fig.~\ref{fig10}(d), both the proposed and Abbas methods provide estimates similar to the theoretical fundamental-mode values over approximately 1--6\,Hz.
The estimates from the Abbas method remain stable up to higher frequencies, particularly for the 500\,m array.
In contrast, as in the isotropic case, the estimates from the proposed method are more stable for the 30\,m array.
For the 30\,m array, the proposed method appears to yield estimates close to the theoretical fundamental-mode attenuation coefficients down to slightly lower frequencies than the Abbas method, whereas no marked difference is observed for the 500\,m array.
One possible reason is that lower power at lower frequencies reduces the SNR, and the resulting loss of estimation accuracy outweighs differences in the applicability of the methods.
Another possible reason is that, as shown in Figs~\ref{fig10}(a) and (b), the phase velocity is almost constant at lower frequencies and the limited variation in wavelength makes the difference between the methods less pronounced.

Figs~\ref{fig10}(a) and (b) also show that, from approximately 8\,Hz, the phase-velocity estimates depart from the fundamental mode and approach the first higher mode, suggesting an increasing contribution from the latter.
Around these frequencies, the attenuation coefficients estimated by the proposed method also become unstable and do not consistently follow the theoretical curve of either mode.
In contrast, the Abbas method with the 500\,m array in the anisotropic case yields values close to the theoretical attenuation coefficients of the first higher mode even at 8\,Hz and above.
This is considered to arise because the proposed method is based on the SPAC method, which cannot separate and identify multiple modes, whereas the Abbas method is based on BFM, which can do so.

\subsection{Detailed examination of wavefield anisotropy}\label{ss:val_mrl_sweep}

\begin{figure*}[tbp]
\centering
\includegraphics[width=\textwidth]{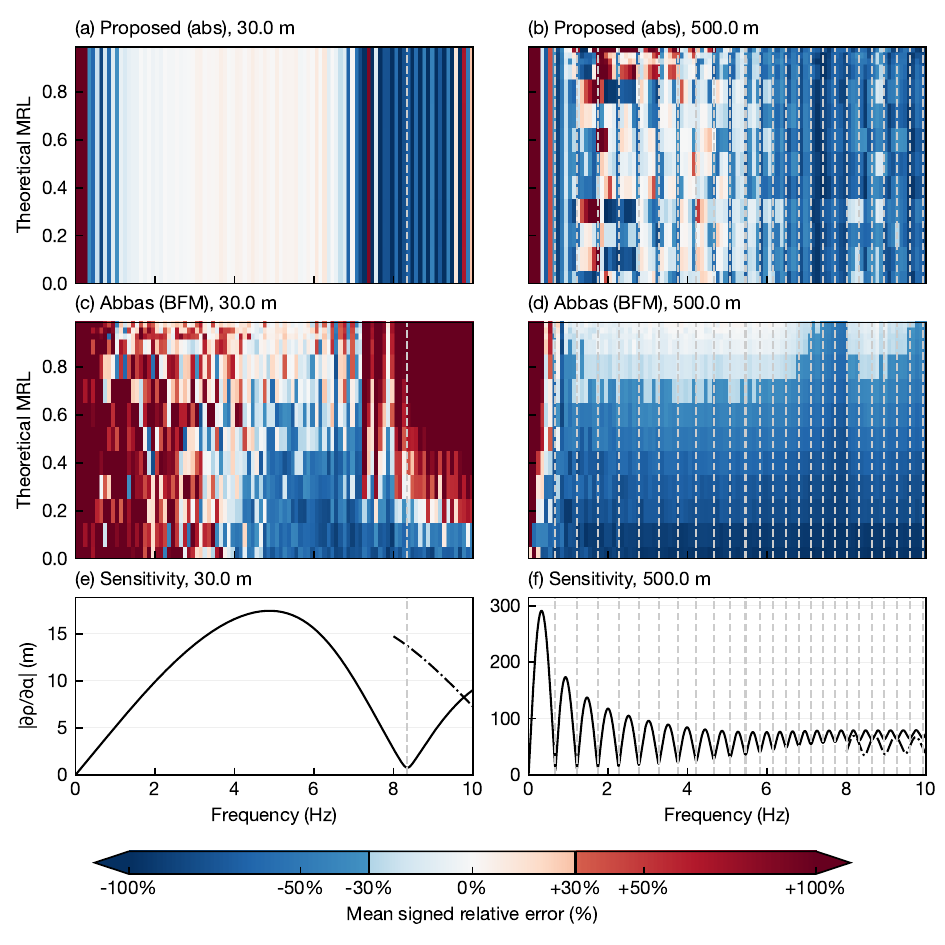}
\caption{Dependence of attenuation estimates on theoretical MRL and array radius. (a, b) Proposed method. (c, d) Abbas method using BFM. Colours indicate signed relative errors averaged over 0.1\,Hz bins, with the fundamental mode used as the reference below 8\,Hz and the first higher mode at and above 8\,Hz. (e, f) Magnitude of the SPAC sensitivity to attenuation for the fundamental (solid) and first higher (dash-dotted) modes. Vertical dashed lines mark the fundamental-mode sensitivity minima.}\label{fig11}
\end{figure*}

Next, the theoretical MRL was varied from 0 to 0.9 in increments of 0.1, with additional values of 0.925, 0.95 and 0.975, to examine the effects of anisotropy in detail.
Fig.~\ref{fig11} shows the signed relative errors of attenuation coefficients averaged over 0.1\,Hz bins.
Considering the change in mode described in the previous subsection, the theoretical attenuation coefficients of the fundamental mode below 8\,Hz and the first higher mode at 8\,Hz and above were used as references.

Figs~\ref{fig11}(a) and (b) show that the proposed method with the 30\,m array gives values close to the theoretical values over 1--6.5\,Hz regardless of MRL, whereas the estimates from the 500\,m array oscillate around the theoretical values.
In both cases, the method fails to estimate the attenuation coefficients of the first higher mode above 6.5\,Hz.
Figs~\ref{fig11}(c) and (d) show that the Abbas method with the 500\,m array gives good results at high MRL values, whereas the estimates from the 30\,m array are unstable even at high MRL values.
This is consistent with the large scatter in the estimates for small arrays reported by \citet{Abbas2025}.
The dependence of both methods on the wavelength-to-array-size ratio and MRL is broadly consistent with the synthetic-data results in Section~\ref{ss:mrl_sweep}.

Figs~\ref{fig11}(e) and (f) show the magnitude of the sensitivity of the SPAC coefficient to the attenuation coefficient, $r|J_1((k+\mathrm{i}\alpha)r)|$.
For the 500\,m array, the sensitivity repeatedly increases and decreases with frequency, whereas for the 30\,m array, the fundamental-mode sensitivity becomes small around 8\,Hz.
Turning to Fig.~\ref{fig11}(b), the frequencies at which the accuracy deteriorates over 1--6\,Hz broadly correspond to the sensitivity minima, consistent with the results in the previous section.
However, for the 30\,m array, the reduction in sensitivity and the increase in the higher-mode contribution occur at similar frequencies, so the instability of the estimates cannot be explained by sensitivity alone.

These results confirmed that the proposed method behaves as expected for wavefields generated numerically under more realistic conditions.
Comparison with the Abbas method, particularly using substantially different array sizes, also confirmed that the results are broadly consistent with the various trends obtained in the previous section.

\section{Preliminary application to field data}\label{s:pre}

\begin{figure*}[tbp]
\centering
\includegraphics[width=\textwidth]{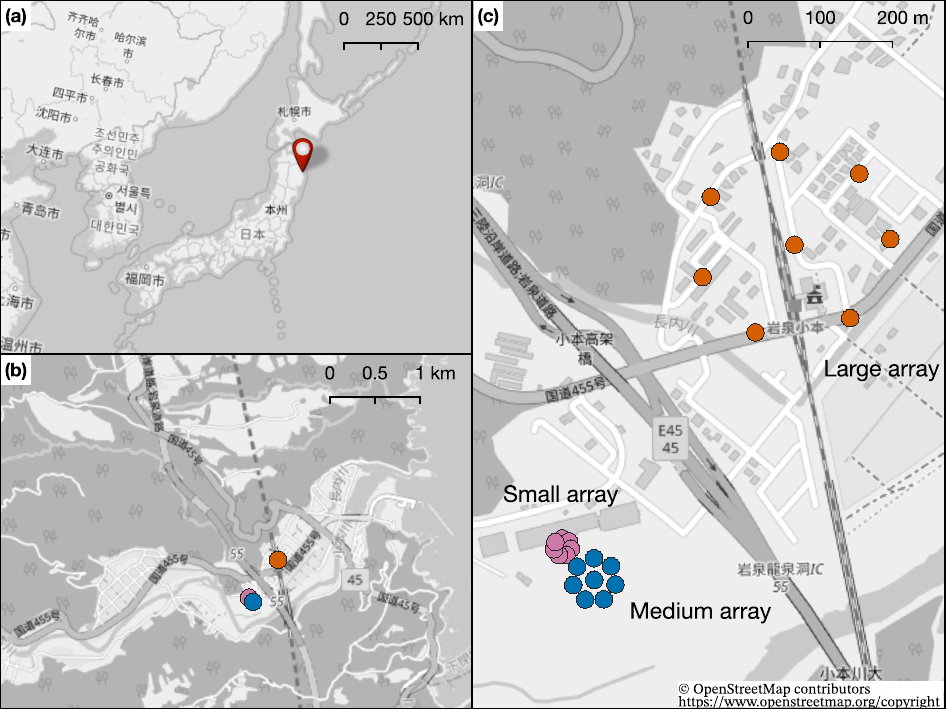}
\caption{Location of the observation site and array configurations in Omoto, Iwaizumi, Iwate, Japan. (a, b) Regional and local maps. (c) Sensor positions of the small, medium and large arrays, with nominal radii of 12, 30 and 132\,m, respectively. Each array consists of seven circumferential sensors and one central sensor.}\label{fig12}
\end{figure*}

To examine the applicability of the proposed method to actual microtremor records, microtremor array measurements were conducted in Omoto, Iwaizumi, Iwate, Japan (Fig.~\ref{fig12}a).
In this section, we examine the consistency of estimates obtained from arrays of different sizes and from the proposed and Abbas methods, and provide a preliminary assessment of their validity by comparison with attenuation coefficients expected from existing empirical relationships.

\subsection{Observation conditions}\label{ss:obs_settings}

Fig.~\ref{fig12} shows the observation site and array configurations.
Measurements were conducted on 19--21 March 2026 using three arrays with radii of 12, 30 and 132\,m, each consisting of 8 observation points: 7 at the vertices of a regular heptagon and 1 at the centre.
The 30\,m array was installed on soil in a school playground, whereas the 12\,m and 132\,m arrays were installed on asphalt pavement.
The 12\,m array used CloverTech LA-352 instruments equipped with Epson M-A352 sensors to record acceleration at 200\,sps.
The 30\,m and 132\,m arrays used Kinkei System KVS-300 moving-coil velocity sensors and aLab AK-002 instruments to record velocity at 100\,sps.
After installation, a step voltage was applied to the velocity sensors using a button cell and the responses were recorded; the natural periods and damping ratios were then identified in post-processing, and instrument corrections were applied to convert the observed records to ground velocity.
The 12 m array recorded for 3 hours from 15:00 on 19 March; the 30 m array for 1 hour from 16:34 to 17:34 on 19 March and 11 hours from 19:00 on 19 March to 05:59 on 20 March (12 hours in total); and the 132 m array for 4.25 hours from 14:45 to 19:00 on 20 March and 12.5 hours from 20:00 on 20 March to 08:30 on 21 March (16.75 hours in total), all in JST.

Segments of 40.96\,s with 50\% overlap were extracted from blocks of the vertical-component records without time-synchronisation or data-storage problems.
Segments were selected using the ratio of the RMS of each segment to that of the entire record, retaining those with ratios near the mode of the histogram.
The bin width was 0.1, and the permitted deviations from the mode were $\pm 0.3$, $\pm 0.15$ and $\pm 0.6$ for the 12, 30 and 132\,m arrays, respectively.
Consequently, 216 (41.1\%), 741 (41.5\%) and 1185 (38.9\%) segments out of totals of 526, 1784 and 3050, respectively, were used for analysis.
The power and cross-spectra were smoothed using a Parzen window with a bandwidth of 0.5\,Hz, and the proposed and Abbas methods were applied to the arrays consisting of the centre point and 7 circumferential points.

\subsection{Initial phase velocities and fit to the SPAC coefficients}\label{ss:pre_fit}

\begin{figure*}[tbp]
\centering
\includegraphics[width=\textwidth]{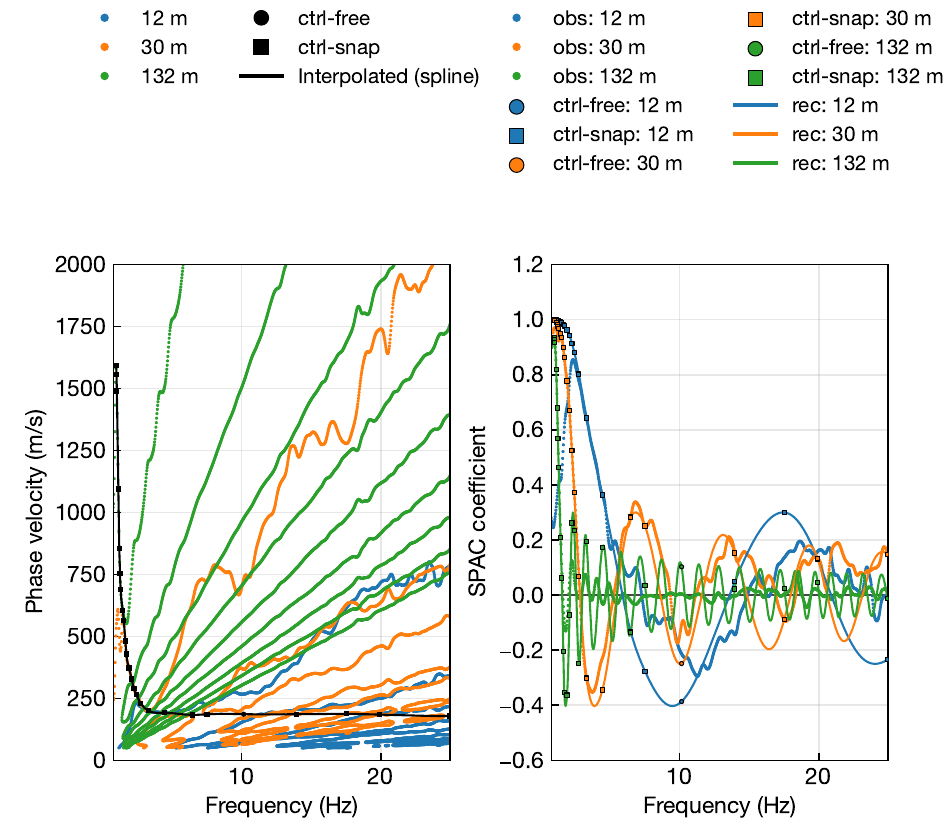}
\caption{Selection of the reference phase-velocity curve. Left: phase-velocity candidates obtained from the real parts of the observed SPAC coefficients and the manually selected, spline-interpolated curve. Right: observed real SPAC coefficients (obs) and elastic SPAC coefficients reconstructed from the selected curve (rec) for the three array radii. Circles (ctrl-free) and squares (ctrl-snap) denote freely placed control points and points snapped to phase-velocity candidates, respectively; their corresponding elastic SPAC values are shown on the right.}\label{fig13}
\end{figure*}

Fig.~\ref{fig13} shows the initial phase velocities determined using all arrays.
Following the method of \citet{Yoshimi2026}, phase velocities were manually picked from multiple candidates obtained from the real parts of the SPAC coefficients, considering the continuity of the dispersion curve and consistency among arrays.
The curve obtained by spline interpolation of the selected points was used as $c_{\mathrm{init}}$, and $c$ and $\alpha$ were estimated using differential evolution following the procedure in Section~\ref{s:method}.
The values of $J_0(\omega r/c_{\mathrm{init}})$ shown on the right of the figure broadly reproduce the main variations in the real parts of the observed SPAC coefficients for each array.
However, at higher frequencies, the amplitudes of the observed values are smaller than those of the reconstructed values, particularly for the 132\,m array.
The following analysis was therefore restricted to the range satisfying both $f<15\,\mathrm{Hz}$ and $k_{\mathrm{ref}}r<16.47$.
Here, $k_{\mathrm{ref}}=\omega/c_{\mathrm{init}}$, and 16.47 corresponds to the third trough of the $J_0$ function.
This upper limit excludes the higher-frequency range with large discrepancies and does not guarantee that directional aliasing is negligible within the retained range.

\begin{figure*}[tbp]
\centering
\includegraphics[width=\textwidth]{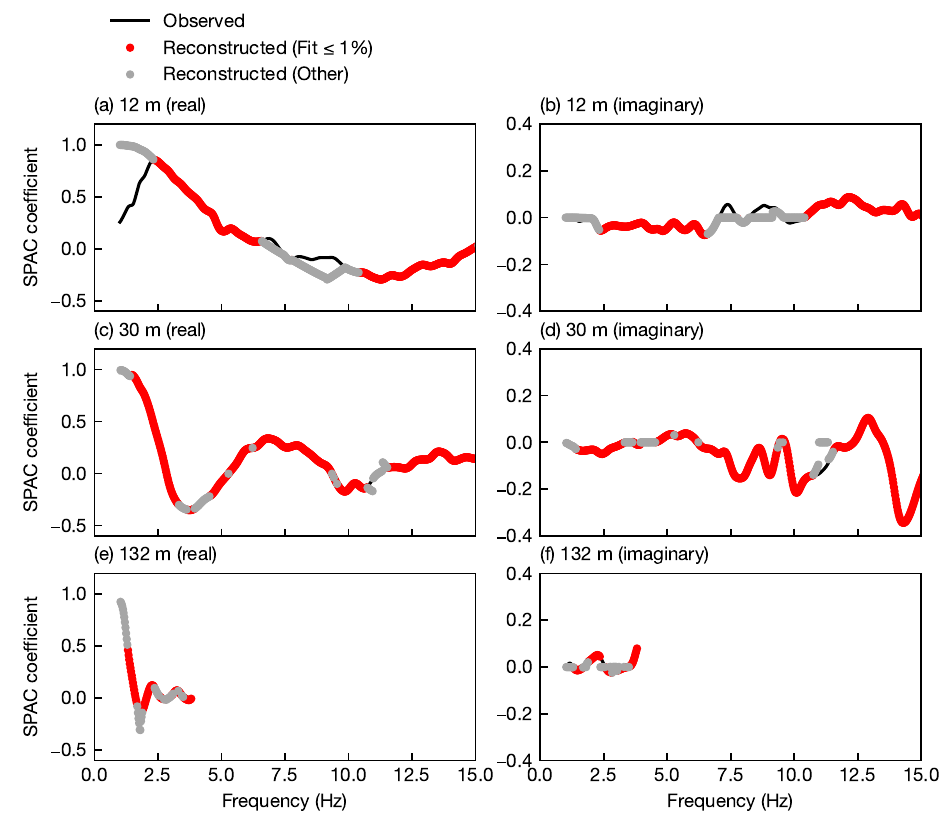}
\caption{Observed and reconstructed SPAC coefficients for the (a, b) 12\,m, (c, d) 30\,m and (e, f) 132\,m arrays. The left and right columns show the real and imaginary parts, respectively. Red symbols indicate frequencies at which both components are reproduced with relative errors of 1\% or less; grey symbols indicate the other estimates.}\label{fig14}
\end{figure*}

Fig.~\ref{fig14} shows the SPAC coefficients calculated from the observed records and those reconstructed from the estimated $(c,\alpha)$ using Eq.~\eqref{e:spac}.
As in Section~\ref{ss:basics}, frequency points at which both the real and imaginary parts have relative reconstruction errors of 1\% or less with respect to the observed values are shown in red.
This does not necessarily mean that the obtained $c$ and $\alpha$ are correct, but it at least indicates the existence of a solution that can simultaneously explain both the real and imaginary parts.
For the 12\,m and 30\,m arrays, both parts are simultaneously reproduced at many frequencies.
For the 132\,m array, many frequency points do not satisfy the fit criterion, although acceptable fits are obtained at some frequencies, mainly at the lower end of the frequency range.

\subsection{Estimation results and validity of the attenuation coefficients}\label{ss:pre_estimates}

\begin{figure*}[tbp]
\centering
\includegraphics[width=\textwidth]{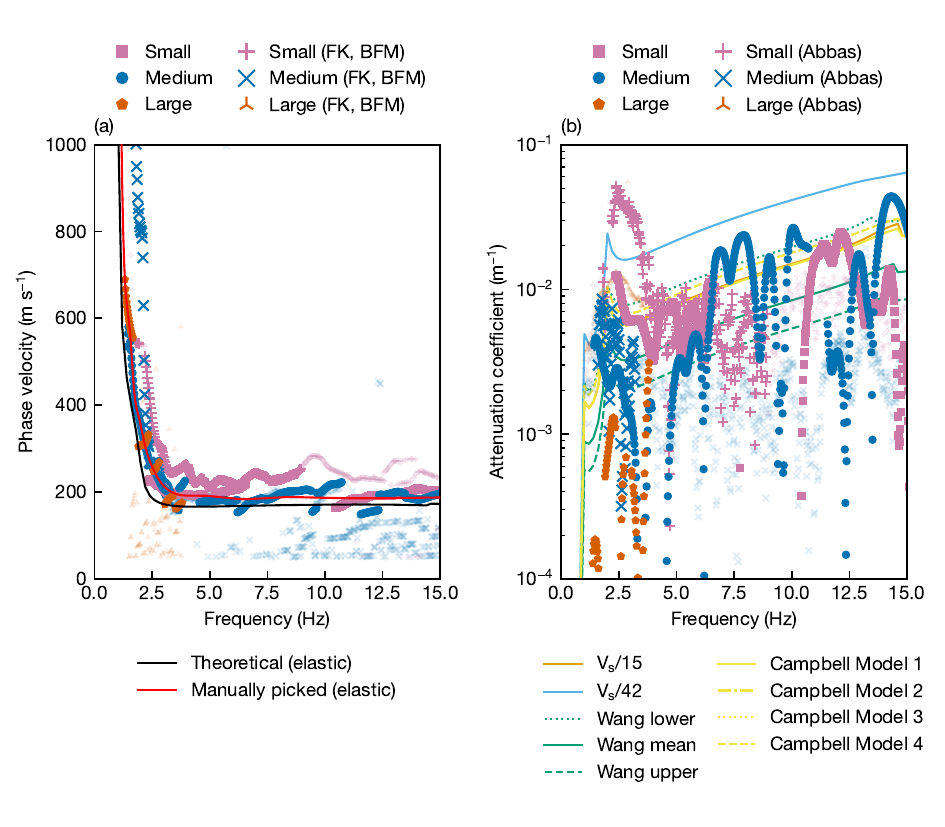}
\caption{Field-data estimates and comparison with reference curves. (a) Phase velocities estimated by the proposed method and conventional beamforming (BFM), together with the manually selected curve and the theoretical elastic curve. (b) Attenuation coefficients estimated by the proposed and Abbas methods, together with theoretical fundamental-mode curves calculated using empirical $Q$ models. Proposed-method estimates satisfy the component-wise 1\% SPAC fit criterion. BFM and Abbas estimates are highlighted where the FK-based MRL exceeds 0.2 and are shown faintly elsewhere.}\label{fig15}
\end{figure*}

Fig.~\ref{fig15} shows the phase velocities and attenuation coefficients that satisfy the reconstruction-error criterion described above.
For comparison, phase velocities obtained by BFM and attenuation coefficients obtained by the Abbas method are also shown.
At frequencies where the MRL calculated from the FK spectrum is below 0.2, the actual MRL is assumed to be correspondingly small.
The Abbas method is therefore assumed not to function correctly at these frequencies, and the estimates from these comparison methods are shown in pale colours.

Fig.~\ref{fig15}(a) shows that the phase velocities estimated by the proposed method are broadly consistent among arrays and do not deviate considerably from the manually picked curve.
The BFM estimates are somewhat larger for the 12\,m array, but for the 30\,m array they are close to those of the proposed method in some frequency ranges.
Fig.~\ref{fig15}(b) shows that the attenuation coefficients estimated by the proposed method are generally of the same order of magnitude for the 12 m and 30 m arrays.
Furthermore, the frequency ranges in which estimation succeeds for these arrays do not overlap greatly, and where estimation fails for one array, the other tends to provide values broadly consistent with the estimates at both ends of that frequency range.
The estimates from the Abbas and proposed methods also appear to be partly consistent, particularly around 5 Hz where the MRL exceeds 0.2.
These results suggest that the attenuation coefficients have been estimated with some degree of reliability.

Furthermore, $Q$ values obtained from existing empirical relationships were assigned to the velocity and density model based on PS and density logging at the site~\cite{Sakai2026}.
Theoretical fundamental-mode attenuation coefficients were then calculated using gRTm.
Specifically, $Q_{\mathrm{S}}$ was calculated using the relationships $Q_{\mathrm{S}}=V_{\mathrm{S}}/15$ and $V_{\mathrm{S}}/42$ compared by \citet{Sato2004}, the mean, upper and lower relationships reported by \citet{Wang1994} and the four models of \citet{Campbell2009}, with $Q_{\mathrm{P}}=Q_{\mathrm{S}}$ in each layer.
Fig.~\ref{fig15}(b) shows that, at 3\,Hz and above, the estimates generally fall within the range indicated by these theoretical curves, except for points with locally small values.
Thus, the attenuation coefficients estimated from actual microtremor records using the proposed method do not appear to substantially contradict existing empirical relationships.

However, the curves based on empirical relationships are neither direct measurements of attenuation at the site nor confidence intervals for the estimates.
Because they also involve extrapolation beyond the applicable ranges of the empirical relationships and the assumption of $Q_{\mathrm{P}}=Q_{\mathrm{S}}$, these results provide a preliminary check of the magnitude of the attenuation coefficients.
A more direct assessment of estimation accuracy requires comparison with attenuation measurements independent of microtremor array measurements.

\section{Conclusion}\label{s:conclusion}

In this study, a method was proposed for estimating Rayleigh-wave attenuation coefficients from microtremor array measurements based on the attenuation-aware formulation of the spatial autocorrelation (SPAC) method~\cite{Kimura2026a}.
The conclusions reached in this study are summarised as follows:
\begin{enumerate}
\item
  Theoretical analysis of directional aliasing showed that attenuation estimation requires denser azimuthal sampling than conventional phase-velocity estimation.
  Based on this result, a regular heptagonal array was adopted for the proposed method.
\item
  A two-stage procedure was proposed in which initial estimates of phase velocity and the attenuation coefficient are obtained using a first-order approximation of the attenuation-aware SPAC coefficient and are then refined by fitting the complex SPAC coefficient.
\item
  Verification using synthetic data confirmed that the proposed method can recover attenuation coefficients under ideal conditions.
  The results also showed that directional aliasing, limited record length and low sensitivity of the SPAC coefficient to attenuation affect attenuation estimation more strongly than phase-velocity estimation.
\item
  Comparison with the Abbas method~\cite{Abbas2025} showed that the two methods have different ranges of applicability.
  The proposed method was less sensitive to wavefield directionality than the Abbas method, although its estimates were affected by directional aliasing at higher frequencies.
  The Abbas method performed particularly well for strongly directional wavefields.
  The proposed method was applicable to longer wavelengths relative to the array size than the Abbas method.
\item
  Results obtained with relative-coordinate arrays were reliable only for nearly unidirectional wavefields, suggesting that their use should generally be avoided when estimating attenuation coefficients.
\item
  Validation using waveforms generated by independent P--SV wave propagation simulations confirmed that the proposed method also functions when the forward model used to generate the wavefield differs from that assumed in the estimation procedure.
\item
  A preliminary application to field microtremor records yielded attenuation coefficients that were broadly consistent across array sizes and, above approximately 3~Hz, with the range expected from existing empirical attenuation relationships.

\end{enumerate}

Further validation against independent attenuation measurements is required to quantify the accuracy of the proposed method. 
Future work should also focus on refining the estimation procedure and relaxing constraints on array configurations.

%
%
\appendix

\section{Relative-coordinate arrays: two extreme cases}\label{app:rca_edge_cases}

Under the assumption of plane-wave propagation, the spectral increment is expected to satisfy the following relationship for an offset $\bm{d}$ with magnitude $d=|\bm{d}|$:
\begin{align}
d\zeta(\bm{x}_0+\bm{d};\omega,k,\phi)
=
\exp\left[
-\mathrm{i}k(\omega)\bm{d}\cdot\bm{p}
\right]
\exp\left[
-\alpha(\omega)\bm{d}\cdot\bm{p}
\right]
d\zeta(\bm{x}_0;\omega,k,\phi)
\end{align}

We now consider using the pair $\bm{x}_0 + \bm{d}$ and $\bm{x}_0  + \bm{r}+ \bm{d}$ in place of the pair $\bm{x}_0$ and $\bm{x}_0 + \bm{r}$.
The autocovariance function $G(\bm{x}_0+\bm{d};\omega,r,\theta)$ is obtained by replacing $\bm{x}_0$ in Eq.~\eqref{e:autocov_fn} with $\bm{x}_0+\bm{d}$.
In that integral, $f(\bm{x}_0;\omega,\phi)$ is then replaced by $f(\bm{x}_0+\bm{d};\omega,\phi)$.
For an infinitesimal region containing a given $(\omega,k,\phi)$, the FWDSD and the spectral increment are related as follows:
\begin{align}
F(\bm{x}_0;\omega,k,\phi)\,\mathrm{d}\omega\,k\mathrm{d}k\,\mathrm{d}\phi
&=E\left[\left|d\zeta(\bm{x}_0;\omega,k,\phi)\right|^2\right]
\end{align}
For an offset $\bm{d}$, the propagation law above gives
\begin{align}
F(\bm{x}_0+\bm{d};\omega,k,\phi)
&=\mathrm{e}^{-2\alpha(\omega)\bm{d}\cdot\bm{p}}
F(\bm{x}_0;\omega,k,\phi)
\end{align}
and hence
\begin{align}
  f(\bm{x}_0 + \bm{d}; \omega, \phi)=\mathrm{e}^{-2\alpha(\omega)\bm{d}\cdot\bm{p}}f(\bm{x}_0; \omega, \phi)
\end{align}
With $\bm{p}'=(\cos\phi',\sin\phi')$, the NFDSD defined in Eq.~\eqref{e:nfdsd} takes the following form at an offset $\bm{d}$:
\begin{align}
\lambda(\bm{x}_0+\bm{d};\omega,\phi)
=
\frac{
\mathrm{e}^{-2\alpha(\omega)\bm{d}\cdot\bm{p}}
\lambda(\bm{x}_0;\omega,\phi)
}{
\displaystyle
\int_{-\pi}^{\pi}
\mathrm{e}^{-2\alpha(\omega)\bm{d}\cdot\bm{p}'}
\lambda(\bm{x}_0;\omega,\phi')
\,d\phi'
}.
\end{align}
Therefore, in a relative-coordinate array, the NFDSD may differ between pairs of observation points, which may cause the SPAC coefficient to deviate from the theoretical value.

We now examine two extreme cases.
The first is a unidirectional wavefield with $\lambda(\bm{x}_0; \omega, \phi)=\delta(\phi - \phi_{\mathrm{s}})$.
With $\bm{p}_{\mathrm{s}}=(\cos\phi_{\mathrm{s}},\sin\phi_{\mathrm{s}})$, we obtain
\begin{align}
\lambda(\bm{x}_0+\bm{d};\omega,\phi)
=
\frac{
\mathrm{e}^{-2\alpha(\omega)\bm{d}\cdot\bm{p}}
\lambda(\bm{x}_0;\omega,\phi)
}{
\mathrm{e}^{-2\alpha(\omega)\bm{d}\cdot\bm{p}_{\mathrm{s}}}
}
=
\lambda(\bm{x}_0; \omega, \phi)
\end{align}
and the NFDSD has no spatial dependence.
This corresponds to the agreement between the RCA and ACA estimates at high MRL values in Fig.~\ref{fig08}.

The second is a perfectly isotropic wavefield.
In this case, $\lambda(\bm{x}_0; \omega, \phi) = \frac{1}{2\pi}$, giving
\begin{align}
\lambda(\bm{x}_0+\bm{d};\omega,\phi)
=
\frac{
\mathrm{e}^{-2\alpha(\omega)\bm{d}\cdot\bm{p}}
\lambda(\bm{x}_0;\omega,\phi)
}{
\frac{1}{2\pi}
\displaystyle
\int_{-\pi}^{\pi}
\mathrm{e}^{-2\alpha(\omega)\bm{d}\cdot\bm{p}'}
\,d\phi'
}
=
\frac{
\mathrm{e}^{-2\alpha(\omega)\bm{d}\cdot\bm{p}}
}{
I_0\left(2\alpha(\omega)d\right)
}
\lambda(\bm{x}_0;\omega,\phi)
\end{align}
The autocorrelation function is then expressed as follows:
\begin{align}
\gamma(\bm{x}_0 + \bm{d}; \omega, r, \theta)
= 
\frac{1}{2\pi I_0\left(2\alpha(\omega)d\right) }\int_{-\pi}^{\pi}
\mathrm{e}^{\mathrm{i}k(\omega) \bm{r}\cdot\bm{p}
-\alpha(\omega)\bm{r}\cdot\bm{p}
-2\alpha(\omega)\bm{d}\cdot\bm{p}}
\mathrm{d}\phi
\end{align}
Denoting the SPAC coefficient approximated using $\gamma(\bm{x}_0 + \bm{d}; \omega, r, \theta)$ by $\rho_{\mathrm{RCA}}$, we obtain
\begin{align}
  \rho_{\mathrm{RCA}}(\omega,r)
=
\frac{1}{(2\pi)^2 I_0\left(2\alpha(\omega)d\right)}
\int_{-\pi}^{\pi}
\int_{-\pi}^{\pi}
\mathrm{e}^{\mathrm{i}k(\omega) \bm{r}\cdot\bm{p}
-\alpha(\omega)\bm{r}\cdot\bm{p}
-2\alpha(\omega)\bm{d}\cdot\bm{p}}
\mathrm{d}\phi
\mathrm{d}\theta
\end{align}
If the offset $\bm{d}$ is constant, the theoretical SPAC coefficient is entirely unaffected.
Here, we consider the case in which $\bm{d}$ depends on $\theta$, as in RCAs often used in the conventional SPAC method.

More specifically, we consider a circular array such as that in Fig.~\ref{fig08}(b) and, for each vertex serving as the starting point of a pair, take $\bm{d}$ to be the translation vector from the centre to that vertex.
With $\bm{d} = (d\cos\psi, d\sin\psi)$, the angles $\psi$ and $\theta$ satisfy $\psi = \Delta + \theta$ for some constant $\Delta$.
Furthermore, $|\bm{r} + \bm{d}|=d$ and $\bm{r}\cdot\bm{d} = rd\cos\Delta= -r^2 / 2 \Leftrightarrow 2d \cos\Delta = -r$.

In this case, setting $\phi'=\phi-\theta$ gives
\begin{align}
\int_{-\pi}^{\pi}
\mathrm{e}^{\mathrm{i}k(\omega) \bm{r}\cdot\bm{p}
-\alpha(\omega)\bm{r}\cdot\bm{p}
-2\alpha(\omega)\bm{d}\cdot\bm{p}}
\mathrm{d}\theta
&=
\int_{-\pi}^{\pi}
\mathrm{e}^{\left\{(\mathrm{i}k - \alpha)r - 2d\alpha\cos\Delta \right\}\cos\phi' - \left(2 d\alpha\sin\Delta \right)\sin\phi'}
\mathrm{d}\phi'
\end{align}
In general, the following identity holds:
\begin{align}
\int_{-\pi}^{\pi} \mathrm{e}^{A\cos u + B \sin u} \mathrm{d} u
=
2\pi I_0 \left(\sqrt{A^2 + B^2}\right)
\end{align}
Thus, setting
\begin{align}
A &= (\mathrm{i}k - \alpha )r - 2d\alpha\cos\Delta =
\mathrm{i}kr\\
B &= -2d\alpha \sin \Delta
\end{align}
gives
\begin{align}
A^2+B^2 
&=-k^2 r^2 + (4d^2 - r^2)\alpha^2
\end{align}

Hence, we obtain
\begin{align}
  \rho_{\mathrm{RCA}}(\omega,r)
&=
\frac{
J_0\left(
\sqrt{
k^2r^2-\alpha^2(4d^2-r^2)
}
\right)
}{
I_0(2\alpha d)
}
\end{align}
where we have assumed $k^2 r^2 > \alpha^2 (4d^2 - r^2)$.

The right-hand side is entirely real.
Furthermore, $I_0$ is an even function, and its Taylor expansion about zero gives $I_0(2\alpha d) \simeq 1 + \alpha^2 d^2$ for $\alpha d \ll 1$.
A first-order approximation assuming $\alpha d \ll 1$ and $\alpha r \ll 1$ gives
$$
\rho_{\mathrm{RCA}}(\omega,r)
\simeq
J_0(kr)
$$
which is the SPAC coefficient in the absence of attenuation.
Therefore, in an isotropic wavefield, attenuation effects appear only at second and higher orders, making it extremely difficult to estimate attenuation using a relative-coordinate array.
This contributes to the marked underestimation in the RCA results for MRL=0 in Fig.~\ref{fig08}.

\section*{CRediT authorship contribution statement}

CRediT: HK: Conceptualization, Data curation, Formal analysis, Funding acquisition, Investigation, Methodology, Project administration, Resources, Software, Visualization, Writing – original draft, Writing – review \& editing; HM: Conceptualization, Data curation, Supervision, Writing – review \& editing.

\section*{Declaration of generative AI use}

During the preparation of this work, the authors used ChatGPT and Codex (OpenAI) to assist with English-language editing and with the development and debugging of analysis and visualisation code.
All AI-assisted content and code were reviewed and validated by the authors, who take full responsibility for the content of the article.

\section*{Declaration of competing interests}

The authors declare that there are no conflicts of interest.

\section*{Acknowledgements}

We gratefully acknowledge Omoto Junior High School and the Omoto Branch Office of Iwaizumi Town for their kind cooperation and support during the field observations in Omoto, Iwate, Japan. 
We also thank Mr Kenta Kobayashi of the Port and Airport Research Institute and Mr Shota Hayashi of the Institute of Science Tokyo for their valuable assistance with the observations.
We are also grateful to Dr Masayuki Yoshimi of the National Institute of Advanced Industrial Science and Technology (AIST) for his valuable advice on the implementation of a tool for manually picking phase velocities from SPAC coefficients.
This work was partially supported by JSPS KAKENHI Grant Number JP25KJ1221.

\section*{Data availability}

The synthetic, simulated and field data analysed in this paper have been deposited in a Mendeley data repository~\cite{Kimura2026data} under the Creative Commons Attribution Non Commercial 4.0 licence.


\bibliographystyle{elsarticle-num-names}
\bibliography{ref.bib}


\end{document}